\documentclass[journal]{IEEEtran}
\usepackage{tipa}
\usepackage{bbm}
\usepackage{mathrsfs}
\usepackage{cite,url,subfigure,epsfig,graphicx}
\usepackage{amssymb,amsmath}
\usepackage{indentfirst}
\usepackage{algorithmic}
\usepackage{algorithm}
\usepackage{pdfpages}
\usepackage{times,verbatim,amsfonts,amsmath,color}
\usepackage{pifont}
\usepackage{multirow}
\usepackage{booktabs}
\usepackage{adjustbox}
\usepackage{hyperref}
\usepackage{threeparttable}
\usepackage{pifont}

\usepackage{booktabs}  
\usepackage{array}    
\usepackage{makecell} 
\usepackage[edges]{forest}
\usepackage{xcolor}

\IEEEoverridecommandlockouts
\makeatletter
\def\ps@headings{\def\@oddhead{\mbox{}\scriptsize\rightmark \hfil \thepage}\def\@evenhead{\scriptsize\thepage \hfil \leftmark\mbox{}}\def\@oddfoot{}\def\@evenfoot{}}
\makeatother 

\newcommand{\tabincell}[2]{\begin{tabular}{@{}#1@{}}#2\end{tabular}}

\usepackage{geometry}
\begin{document}
\title{Internet of Agents: Fundamentals, Applications, and Challenges}
\author{Yuntao~Wang, Shaolong~Guo, Yanghe~Pan, Zhou~Su, Fahao~Chen, Tom H.~Luan, Peng~Li, Jiawen~Kang, and Dusit~Niyato 
\thanks{Y.~Wang, S.~Guo, Y.~Pan, Z.~Su, T.~H.~Luan, and P.~Li are with the School of Cyber Science and Engineering, Xi'an Jiaotong University, Xi'an, China. \textit{(Corresponding author: Zhou~Su)}}
\thanks{F.~Chen is with the School of Artificial Intelligence, Shandong University, China.}
\thanks{J.~Kang is with the School of Automation, Guangdong University of Technology, Guangzhou, China.}
\thanks{D.~Niyato is with the College of Computing and Data Science, Nanyang Technological University, Singapore.}
\thanks{Accepted by IEEE TCCN in Oct. 2025. }
}

\maketitle

\begin{abstract}
With the rapid proliferation of large language models and vision‑language models, AI agents have evolved from isolated, task‑specific systems into autonomous, interactive entities capable of perceiving, reasoning, and acting without human intervention. As these agents proliferate across virtual and physical environments, from virtual assistants to embodied robots, the need for a unified, agent‑centric infrastructure becomes paramount. In this survey, we introduce the Internet of Agents (IoA) as a foundational framework that enables seamless interconnection, dynamic discovery, and collaborative orchestration among heterogeneous agents at scale. We begin by presenting a general IoA architecture, highlighting its hierarchical organization, distinguishing features relative to the traditional Internet, and emerging applications. Next, we analyze the key operational enablers of IoA, including capability notification and discovery, adaptive communication protocols, dynamic task matching, consensus and conflict‑resolution mechanisms, and incentive models.  
Finally, we identify open research directions toward building resilient and trustworthy IoA ecosystems.
\end{abstract}


\IEEEpeerreviewmaketitle
\section{Introduction}
The rapid advancement of large models, including large language models (LLMs) and vision-language models (VLMs), has ushered in a new era of artificial intelligence (AI) agents (or agentic AI) \cite{xi2023rise,wang2024large}, transforming them from isolated, task-specific models into autonomous, interactive entities. 
These agents can perceive, reason, and act independently, capable of seamless collaboration with humans and other agents in complex environments, marking a pivotal step toward artificial general intelligence (AGI) \cite{cheng2024exploring,zhao2024see}.
From virtual assistants to physically embodied systems such as humanoid robots, autonomous unmanned aerial vehicles (UAVs), and intelligent vehicles, AI agents are rapidly weaving into daily life \cite{chen2024persona}. 
Tech giants have declared their adventure to develop next-generation AI agents, such as OpenAI Operator and ByteDance's Doubao Agent.
For instance, platforms such as Hugging Face host over 1 million open-source models, while Tencent Yuanbao support over 100K specialized agents. According to Gartner \cite{GartnerAgent}, by 2028, at least 15\% of daily tasks will be autonomously performed by AI agents, and 33\% of enterprise applications will incorporate agent-driven intelligence. 
As AI agents proliferate, they are poised to act as ``new citizens" in digital and physical spaces, reshaping economic structures and human social interactions.

The widespread adoption of agents has spurred the need for real-time cross-domain agent communication and coordination, particularly in scenarios such as smart cities including millions of heterogeneous agents. 
The Internet of agents (IoA) \cite{chen2025ioa,aminiranjbar2024dawn} 
emerges as a foundational infrastructure for next-generation intelligent systems that enables seamless interconnection, autonomous agent discovery, dynamic task orchestration, and collaborative reasoning among large-scale virtual/embodied agents.
Unlike the human-centric Internet, IoA is agent-centric and prioritizes inter-agent interactions \cite{A2A}, where the exchanged information shifts from human-oriented data (e.g., text, images, and audio) to machine-oriented data objects (e.g., model parameters, encrypted tokens, and latent representations). Furthermore, interaction methods are evolving beyond graphical user interfaces (GUIs) toward semantic-aware and goal-driven communications \cite{10798108} via auto-negotiation \cite{Agora}. 
Additionally, by offering scalable networked AI inference and shared sensing capabilities, IoA empowers resource-constrained agents, e.g., mobile devices and UAVs, with access to advanced AI capabilities \cite{qu2024mobile} and beyond-line-of-sight (BLOS) perception.
By fostering seamless interoperability between virtual and embodied agents from distinct entities, IoA introduces new connection paradigms and traffic patterns beyond the current Internet and mobile Internet. As this ecosystem continues to evolve, IoA is poised to become the backbone of a new era of human-AI symbiosis, reshaping global connectivity and pushing the frontiers of intelligent systems. 

\begin{table*}[!t]
\begin{center}\setlength{\abovecaptionskip}{0cm}
	\caption{Summary of Key Abbreviations in Alphabetical Order}\label{table:abbr}
		\begin{tabular}{ll|ll|ll}
			\toprule
			\textbf{Abbr.}  &\textbf{Definition}            & \textbf{Abbr.} &\textbf{Definition}           &\textbf{Abbr.} &\textbf{Definition} \\
			\midrule
A2A   & Agent-to-Agent       & AGI   & Artificial General Intelligence & AI  & Artificial Intelligence \\       
ANP   & Agent Network Protocol & BLOS  & Beyond-Line-of-Sight          & CoT & Chain-of-Thought  \\                 
CUA   & Computer-Use Agent   & D2D   & Device-to-Device                & DID & Decentralized IDentifier \\
DNN   & Deep Neural Network  & EOM   & Economy of Minds                & GoT & Graph-of-Thought \\
GUI   & Graphical User Interface & IoA & Internet of Agents            & IoT & Internet of Things  \\
LLM   & Large Language Model & MARL  & Multi-Agent Reinforcement Learning & MAS & Multi-Agent Systems \\
MCP   & Model Context Protocol & MPC  & Multi-Party Computation        & NLP & Natural-Language Processing \\
P2P   & Peer-to-Peer         & PMU   & Phasor Measurement Unit        & pub/sub & Publish-Subscribe \\
QoS   & Quality-of-Service   & RAG   & Retrieval-Augmented Generation & RL  & Reinforcement Learning \\
SSE   & Server-Sent Events   & ToT   & Tree-of-Thought                & UAV & Unmanned Aerial Vehicle \\
UUID  & Universally Unique IDentifier & VC  & Verifiable Credential     & VLM & Vision-Language Model \\
			\bottomrule
		\end{tabular}
	\end{center}
\end{table*}

Despite its bright future, the practical deployment of IoA at scale faces several critical challenges. 
\begin{itemize}
    \item \textit{Interconnectivity:} Existing multi-agent systems (MAS) are primarily simulated on single devices, whereas real-world IoA deployments span billions of geographically distributed agents, each with unique compute, network, sensing, and energy profile \cite{chen2025ioa}. This shift necessitates new agent networking architectures that support seamless interoperability among heterogeneous agents, breaking down data silos. Moreover, IoA should provide large models with complete contextual awareness and access to all Internet-based tools \cite{MCP}.
    \item \textit{Agent-Native Interface:} Current computer-use agents (e.g., OpenAI's Operator) rely on mimicking human GUI actions (e.g., clicks and keystrokes) to control browsers and apps \cite{sager2025ai}, incurring high screen-scraping overhead. IoA should empower agents to interact natively (e.g., APIs or semantic communication protocols) with other agents and Internet resources \cite{A2A,MCP}, rather than emulating human behavior. 
    \item \textit{Autonomous Collaboration:} IoA encompasses both physical and virtual agents operating in highly dynamic settings. Embodied agents (e.g., autonomous robots and UAVs) exhibit spatial mobility, while software agents can be instantiated, migrated, or terminated on demand. By harnessing the power of large models, IoA should let agents self-organize, self-negotiate, and form low-cost, high-efficiency collaborative networks for autonomous agent discovery, capability sharing, task orchestration, and load balancing \cite{chen2025ioa}.
\end{itemize}

As IoA continues to evolve, tackling these challenges are essential to realize large-scale IoA deployments and unlock the full potential of next-generation autonomous AI systems.

\subsection{Comparison with Existing Surveys and Contributions of Our Work}\label{subsec:Contributions}

Recent research in large model-based agents and MAS has attracted considerable attention across both academia and industry. 
Jin \emph{et al.} \cite{jin2025comprehensive} review intelligent decision-making approaches, algorithms, and models in MAS and categorize them into rule-based, game theoretical, evolutionary algorithm-based, multi-agent reinforcement learning (MARL)-based, and LLM-based methods. 
Guo \emph{et al.} \cite{guo2024large} systematically examine the development of LLM-based MAS by addressing the agent-environment interface, LLM agent characterization, inter-agent communication strategies, and capability acquisition paradigms, while also highlighting applications in problem-solving and world simulation. 
Tran \emph{et al.} \cite{tran2025multi} classify LLM-based multi-agent collaborative systems based on key characteristics including type, strategy, structure, and coordination. 
Li \emph{et al.} \cite{li2024survey} offer a comprehensive survey on the construction of LLM-based MAS with a focus on problem-solving and world simulation. 
Wang \emph{et al.} \cite{wang2024large} review the overall framework of LLM agents, covering enabling technologies, core characteristics, collaboration paradigms, and their associated security and privacy challenges, along with potential defense strategies in future AI agent systems. 
Wu \emph{et al.} \cite{wu2025multi} present a review of LLM-based multi-agent autonomous driving systems by addressing multivehicle interactions, vehicle-infrastructure communication, and human-vehicle codriving. 
He \emph{et al.} \cite{he2024llm} systematically assess the capabilities and limitations of LLM-based MAS applications in software engineering. 
Amirkhani \emph{et al.} \cite{amirkhani2022consensus} provide an overview of consensus in MAS by reviewing taxonomies, dynamic models, protocols, control mechanisms, and applications.

Existing surveys mainly focus on MAS, which have three challenges. 1) Ecosystem isolation: Existing frameworks limit agents only to their own environments, restricting the integration of third-party agents and reducing the diversity of capabilities. 2) Single-device simulation: Most MAS are confined to single-device simulations, which contrast sharply with real-world scenarios where agents operate across multiple devices and varied geographic locations. 3) Rigid communication and coordination: Most of existing agent protocols and state transitions are hard-coded and fail to capture the dynamic, task-specific nature of practical collaboration.
In contrast, this survey focuses on the networking aspects of large model-based agents, addressing their architectures and open challenges of the Internet of agents (IoA). 
Table~\ref{contribution} compares the contributions of our survey with previous related surveys in the field of IoA. 

\begin{table}[!t]
   \centering \setlength{\abovecaptionskip}{0cm}
    \caption{A Comparison of Our Survey with Relevant Surveys}\label{contribution}
    \resizebox{1\linewidth}{!}{
        \begin{tabular}{c|c|l}
        \toprule
        \textbf{Year.} &\textbf{Ref.} &\textbf{Contribution} \\ \hline 

        {2022} &\cite{amirkhani2022consensus} &\tabincell{l}{Overview of consensus in MAS including taxonomies,\\ models, protocols, control mechanisms, and applications.} \\ \hline
        
        {2024} &\cite{he2024llm} &\tabincell{l}{Discussions on capabilities and limitations of LLM-based \\MAS applications in software engineering.} \\ \hline

        {2024} &\cite{guo2024large} &\tabincell{l}{Survey of LLM-based MAS, including agent-environment \\interface, LLM agent characterization, inter-agent comm., \\capability acquisition, and applications.} \\ \hline
        
        {2024} &\cite{li2024survey} &\tabincell{l}{Survey on LLM-based MAS construction in problem-solving \\and world simulation.} \\ \hline

        {2024} &\cite{wang2024large} &\tabincell{l}{Review enabling technologies, core features, cooperation \\paradigms, security/privacy challenges, and defense\\strategies of LLM agents.} \\ \hline
        
        {2025} &\cite{jin2025comprehensive} &\tabincell{l}{Review intelligent decision-making approaches, algorithms, \\and models in MAS.} \\ \hline
        
        {2025} &\cite{tran2025multi} &\tabincell{l}{Discussions on key characteristics including type, strategy,\\ structure, and coordination of LLM-based MAS.} \\ \hline

        {2025} &\cite{wu2025multi} &\tabincell{l}{Survey on LLM-based multi-agent autonomous driving\\systems including multivehicle interaction, vehicle-infra. \\communication, and human-vehicle codriving. } \\ \hline

        {Now} &\textbf{Ours} &\tabincell{l}{Comprehensive survey of fundamentals, applications, and\\ challenges of IoA, discussions on architecture design, key\\ characteristics, working paradigms, and open issues in IoA.} \\ \bottomrule
        \end{tabular}}
\end{table}

\begin{figure}[!htbp]
\centering \setlength{\abovecaptionskip}{-0.1cm}
\includegraphics[width=0.9\linewidth]{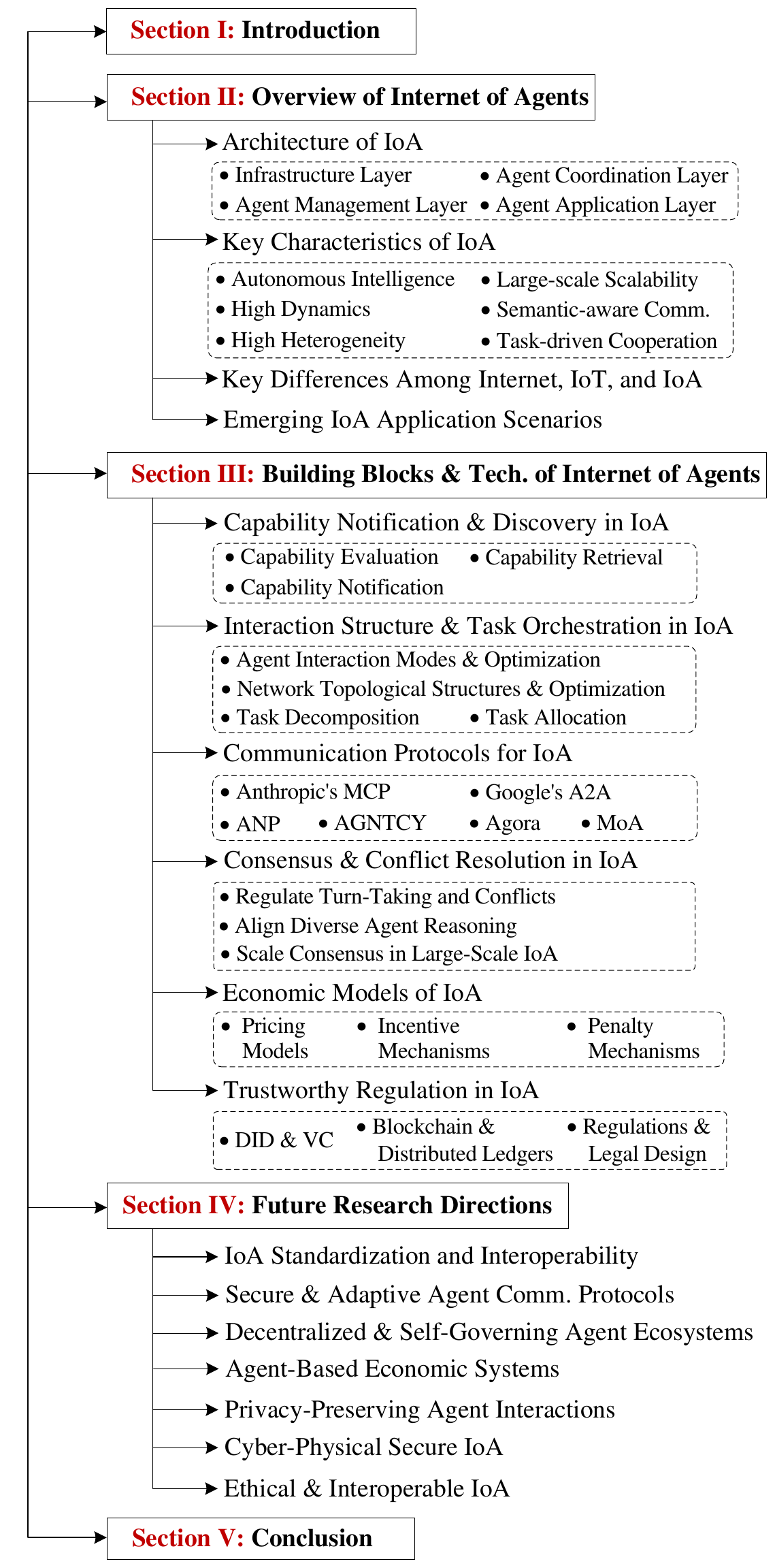}
 \caption{Organization structure of this survey paper.}\label{fig:organization}\vspace{-4mm}
\end{figure}

We provide a comprehensive survey on the fundamentals, emerging applications, and open challenges of IoA, paving the way for its {future development}. 
Our survey aims to: 1) provide a comprehensive understanding of the IoA architecture and its networking mechanisms that facilitate the integration of diverse, distributed agents; and 2) critically examine the operational paradigms of IoA, including dynamic agent discovery, task orchestration, communication protocols, and consensus mechanisms. The main contributions include: 
{\begin{itemize}
    \item \textit{General IoA Architecture.} We provide a holistic overview of the IoA architecture, outlining its hierarchical framework, distinguishing features from the traditional Internet, key characteristics, and emerging applications.
    \item \textit{Key Enablers of IoA.} We analyze critical enablers that support IoA, including capability notification \& discovery, dynamic task orchestration \& matching, adaptive communication protocols, consensus \& conflict resolution mechanisms, economic models, and regulation models.
    \item \textit{Open Challenges of IoA.} We identify unresolved issues in the IoA domain and point out future research directions to foster its broad adoption and sustainable evolution.
\end{itemize}}

\subsection{Paper Organization}\label{subsec:organization}
The remainder of this paper is organized as follows. In Section~\ref{sec:OVERVIEW}, we provide a detailed examination of the IoA architecture. Section~\ref{sec:Paradigms} explores the building blocks and enabling technologies of IoA. 
Finally, Section~\ref{sec:FUTUREWORK} outlines future research directions for advancing IoA. The organizational structure of this survey is illustrated in Fig.~\ref{fig:organization}, and the key acronyms are summarized in Table~\ref{table:abbr}.

\section{Overview of Internet of Agents}\label{sec:OVERVIEW}
In this section, we first introduce the architecture of IoA. We then explore the core characteristics of the IoA paradigm, followed by examining the key differences among the IoA, the traditional Internet, and the Internet of things (IoT). Finally, we highlight emerging applications in IoA across various domains.

\subsection{Architecture of IoA}\label{subsec:Architecture}
The IoA is an emerging agent-centric infrastructure that connects billions of autonomous agents, both virtual and embodied, across diverse domains. It enables real-time communication, discovery, and coordination among heterogeneous agents, supporting complex task orchestration in dynamic environments such as smart cities. 
The advent of the IoA, or agentic Web, represents a fundamental transformation in how users and services engage with the digital realm. 

\textit{1) Agent Types.} {Unlike traditional symbolic agents that rely on rule-based or logic-driven mechanisms for decision-making, LLM-powered agents employ generative models for reasoning, planning, and natural language interaction. In addition, robotic or embodied agents operate in the physical world by coupling perception with actuation. Table~\ref{tab:llmagent_vs_embodied} summarizes the typology of agents. Overall, within the IoA,} agents can be broadly classified into two principal categories: \textit{virtual agents} and \textit{embodied agents}. 
\begin{itemize}
    \item Virtual agents operate entirely within digital environments, including chatbots, virtual assistants, customer-service agents, and other software-based autonomous systems \cite{park2023generative}. They leverage high-bandwidth, wired or persistent network connections to process large-scale language models, access remote knowledge bases, and interact with users via graphical, voice, or wearable interfaces. 
    \item Embodied agents, such as home robots, UAVs, and autonomous vehicles, inhabit the physical world and rely on onboard sensors (e.g., cameras, LiDAR, inertial measurement units) and actuators to perceive and manipulate their environment \cite{zhao2024see}. They generally operate under variable wireless conditions for remote coordination.
\end{itemize}

\begin{table*}[ht]
\centering
\caption{{Comparison Among Symbolic Agents, LLM-powered Virtual Agents, and Robotic/Embodied Agents}}\label{tab:llmagent_vs_embodied}
\begin{tabular}{p{2.2cm}|p{3.5cm}|p{3.4cm}|p{3.15cm}|p{2.7cm}}
\toprule
\textbf{Agent Type} & \textbf{Core Mechanism} & \textbf{Strengths} & \textbf{Limitations} & \textbf{Roles in IoA} \\ \hline

\textbf{Symbolic Agents} & \makecell[l]{Logic-based or rule-driven \\inference and decision-making} 
& \makecell[l]{High interpretability, \\formal guarantees, \\efficient in structured domains} 
& \makecell[l]{Limited scalability, \\less flexible in \\unstructured environments} 
& \makecell[l]{Protocol enforcement, \\rule-based coordination, \\compliance checking} \\ \hline

\makecell[l]{\textbf{LLM-powered}\\ \textbf{Virtual Agents}} & \makecell[l]{Generative large models for \\reasoning, planning, and \\natural language interaction} 
& \makecell[l]{Strong adaptability, \\contextual reasoning, \\multi-domain capability} 
& \makecell[l]{Computationally expensive, \\unpredictable behavior, \\reliance on data quality} 
& \makecell[l]{Conversational agents, \\code generation, \\research assistants} \\ \hline

\makecell[l]{\textbf{Robotic/Embodied}\\ \textbf{Agents}} & \makecell[l]{Coupling perception with \\actuation in the physical world} 
& \makecell[l]{Direct real-world interaction, \\physical task execution} 
& \makecell[l]{Hardware constraints, \\safety/reliability concerns, \\limited reasoning ability} 
& \makecell[l]{Manufacturing, \\autonomous vehicles, \\healthcare robotics} \\ \bottomrule
\end{tabular}
\end{table*}

\begin{figure}[!t]
\centering \setlength{\abovecaptionskip}{-0.cm}
  \includegraphics[width=0.9\linewidth]{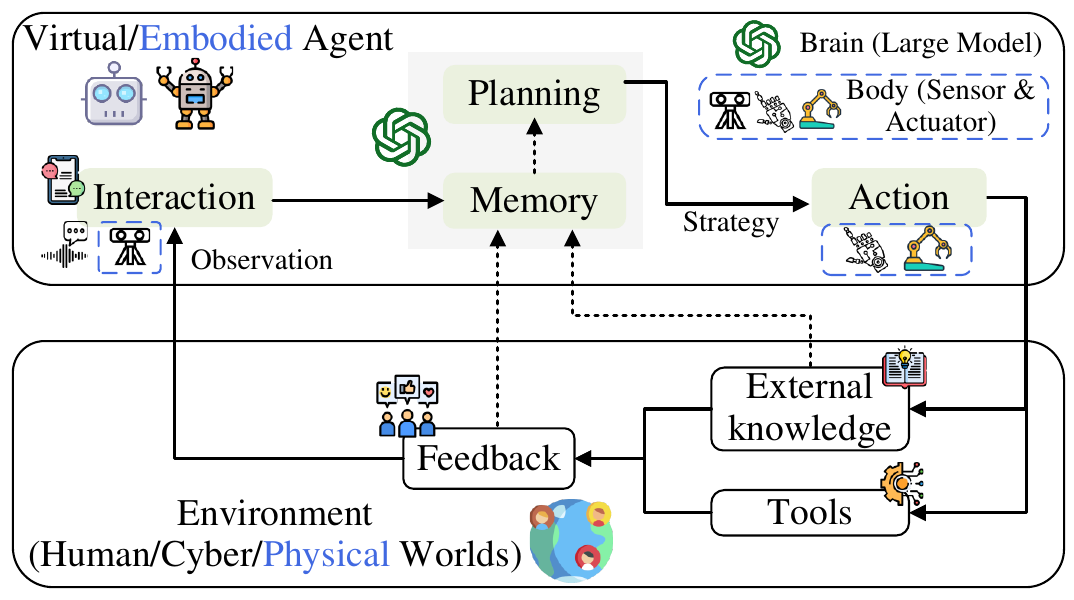}
  \caption{Workflow of functional modules of virtual and embodied agents. Note that the blue text and blue dashed box are unique for embodied agents.}\label{fig:AgentWorkflow}\vspace{-3mm}
\end{figure}

\textit{2) Functional Modules of Agents.} Regardless of form, as depicted in Fig.~\ref{fig:AgentWorkflow}, each IoA agent typically comprises four core functional modules \cite{xi2023rise,wang2024large}: \textit{planning}, \textit{memory}, \textit{interaction}, and \textit{action}, which together empower agents to operate, learn, and adapt in complex environments. 
\begin{itemize}
    \item \textit{Planning.} Fueled by large models, the planning module serves as the brain of the agent and enables advanced reasoning, task decomposition, and decision-making. It continuously consults the memory module to incorporate past experiences and external knowledge, producing context-aware action plans. Typical LLM-driven planning techniques include two modes. (i) \textit{Feedback-free planning:} Sequential and structured reasoning methods that decompose complex tasks into subgoals, such as chain-of-thought (CoT) \cite{wei2023cot}, tree-of-thought (ToT) \cite{yao2023tot}, graph-of-thought (GoT) \cite{besta2024graph}. (ii) \textit{Feedback-enhanced planning:} Integrated reasoning-action frameworks that merge plan generation with task feedback (e.g., real-time execution and self-reflection) to iteratively correct errors and refine strategies, such as ReAct \cite{yao2023ReAct}, Reflexion \cite{shinn2023Reflexion}, and VOYAGER \cite{wang2023voyager}.
    
    \item \textit{Memory.} The memory module underpins an agent's ability to learn, adapt, and personalize its behavior by maintaining both short-term context (e.g., ongoing dialogues and recent task results) and long-term knowledge (e.g., accumulated domain facts and past interaction records) \cite{cheng2024exploring}. By combining real-time updates with historical data, it enables continual learning, personalization, and informed decision-making. Memory can be organized into three types. (i) \textit{Short-term memory} \cite{kang2024knowledge} is a limited context buffer for recent observations and dialogue turns, ensuring coherent, up-to-date reasoning. (ii) \textit{Long-term memory} \cite{trivedi2022interleaving} is external vector stores or knowledge bases that archive historical experiences, domain facts, and policies, accessed via retrieval-augmented generation (RAG) \cite{lewis2020retrieval} for informed planning. (iii) \textit{Hybrid memory} \cite{liu2024RAISE} is a dynamic cache that promotes key short-term fragments into long-term storage and retrieves relevant long-term entries back into the working buffer.
    
    \item \textit{Interaction.} The interaction module enables IoA agents to communicate and collaborate across human, agentic, and environmental interfaces by fusing multimodal inputs, ranging from natural language and emotional cues to sensor streams such as LiDAR and camera, into unified semantic representations. It extracts intent and situational context, enabling agents to maintain awareness of both its physical surroundings and its digital environment. 
    (i) \textit{Agent-agent interactions:} Agents exchange structured knowledge, negotiate roles, assign teamwork, and coordinate shared tasks via semantic protocols. 
    (ii) \textit{Agent-human interactions:} By parsing natural language, affective signals, and dialogue history, agents maintain consistent personas and emotional rapport. Tree-structured persona models \cite{jinxin2023CGMI} help manage character continuity and role assignments over extended conversational exchanges. 
    (iii) \textit{Agent-environment interactions:} Agents engage with physical and virtual surroundings through closed-loop feedback. Techniques such as self-sampling reinforcement learning (RL) and curriculum fine-tuning \cite{Lai2024AutoWebGLM} enable robust performance in web navigation, mixed-reality tasks, and real-world robotics applications.
    
    \item \textit{Action.} The action module brings an agent's plans to life by interfacing with both physical actuators and software tools. It also monitors execution outcomes via the perception module, adjusting behaviors to handle unexpected obstacles or changing objectives. It encompasses two main capabilities: (i) \textit{Embodied operations:} translates high-level plans into real-world behaviors such as robotic grasping, navigation, or manipulation. For instance, in SayCan \cite{ichter2022do}, a warehouse robot can use an LLM to suggest pickup poses (``Say'') and a learned affordance model to verify feasibility (``Can''), ensuring reliable execution under changing conditions. (ii) \textit{Tool invocation \& creation:} empowers agents to call external services such as search engines, APIs, or databases to gather information or execute subtasks. Agents can also synthesize new tools (e.g., custom data‐processing scripts) to expand their action repertoire for improved efficiency.
\end{itemize}

\begin{figure}[!t]
\centering \setlength{\abovecaptionskip}{-0.cm}
\includegraphics[width= 1.03\linewidth]{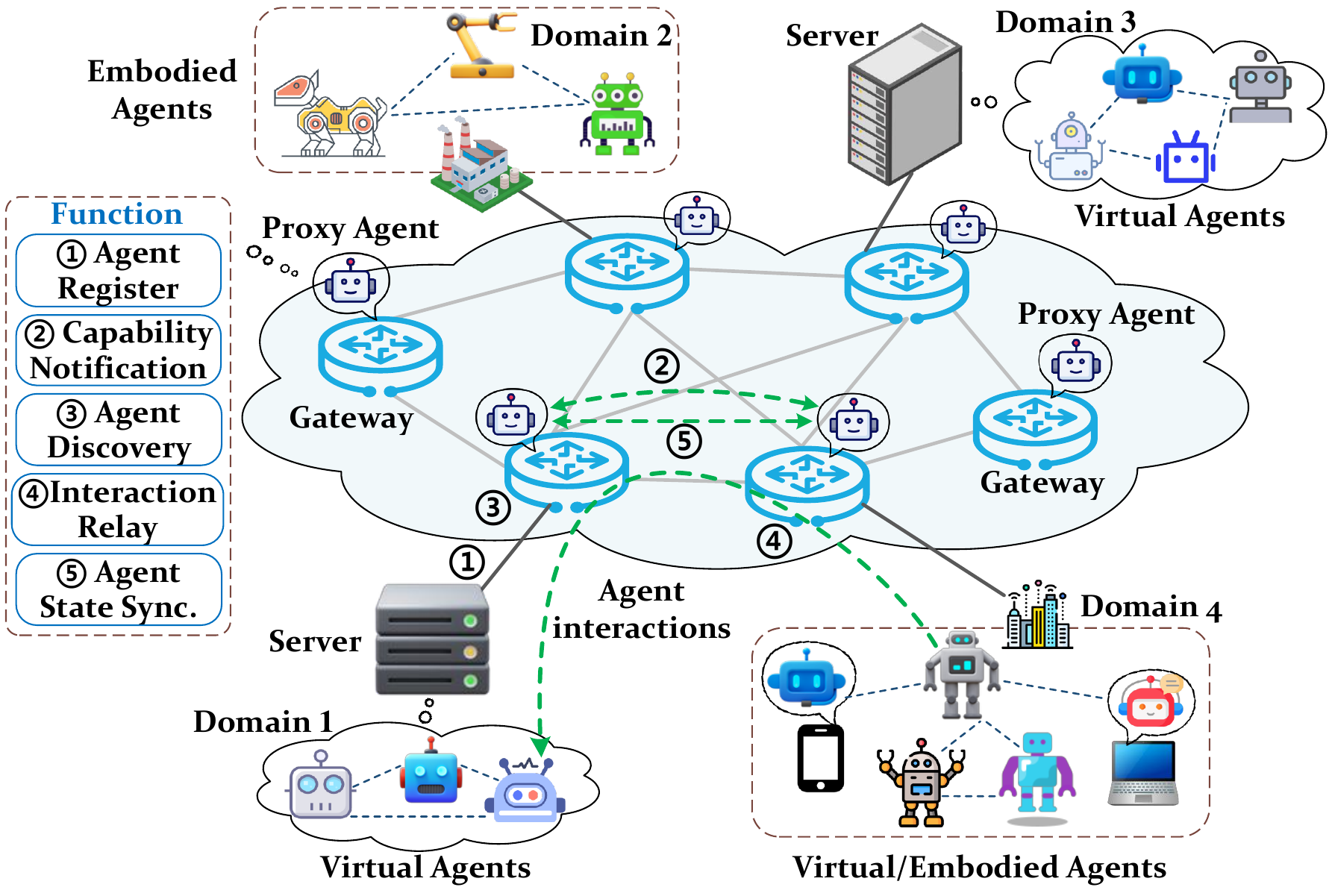}
 \caption{An overview of IoA encompassing heterogeneous autonomous agents across diverse domains, with an illustrative example of cross-domain agent interactions. Consider two agents operating in distinct domains, e.g., domains 1 and 4. Initially, each agent registers with a proxy agent hosted on its respective domain gateway to facilitate agent discovery. Gateway proxy agents then broadcast the capabilities of all registered agents within their domain. Upon identifying a common task, the two agents initiate collaboration, facilitated by relay gateways that manage inter-domain communication. Throughout the interaction, gateway proxy agents coordinate and synchronize the agents' states, ensuring consistency and coherence across domains.}\label{fig:overview}\vspace{-2mm}
\end{figure}

Collectively, these modules empower agents to function autonomously and cooperatively within complex, dynamic environments. Generally, each agent has two parts \cite{wang2024large}: a large model-powered \textit{brain} and a physical or digital \textit{body} (e.g., UAV, robot dog, and digital human). Specifically, based on generality, large models fall into two primary categories: \textit{(i) foundational models} (e.g., DeepSeek-R1 and GPT-4o), which provide general-purpose cognitive capabilities; and \textit{(ii) task-specific large models}, which are fine-tuned for specialized downstream tasks within particular domains. 

\textit{3) Interconnected Sub-IoA.}
As illustrated in Fig.~\ref{fig:overview}, the IoA architecture comprises multiple interconnected and domain-specific agent networks, referred to as sub-IoAs. Each sub-IoA operates semi-autonomously and is anchored by a designated gateway node that hosts a \textit{gateway proxy agent} \cite{aminiranjbar2024dawn}, responsible for orchestrating resources, agent register \& discovery, agent capacity notification within its domain, as well as cross-domain relays and state
synchronization. This hierarchical structure enables scalable, modular deployment across diverse environments. 

To support distributed and efficient IoA management at scale, {gateway proxy agents} \cite{aminiranjbar2024dawn} act as intelligent intermediaries between domain-level task requests and globally distributed resources (e.g., tools, agents, and agentic applications). It provides two principal services: (i) a resource registry for enrollment and discovery, and (ii) an intelligent match-and-retrieve mechanism that semantically interprets task queries and identifies relevant resources accordingly. Upon receiving task requests, the proxy agent searches its local registry, domain-specific filters and policy-based guardrails to rank candidate agents, and returns the most suitable matches. Security measures such as access control policies and automated resource validation further safeguard the system reliability and trustworthiness.

\begin{figure}[!t]
\centering \setlength{\abovecaptionskip}{-0.cm}
\includegraphics[width= 0.92\linewidth]{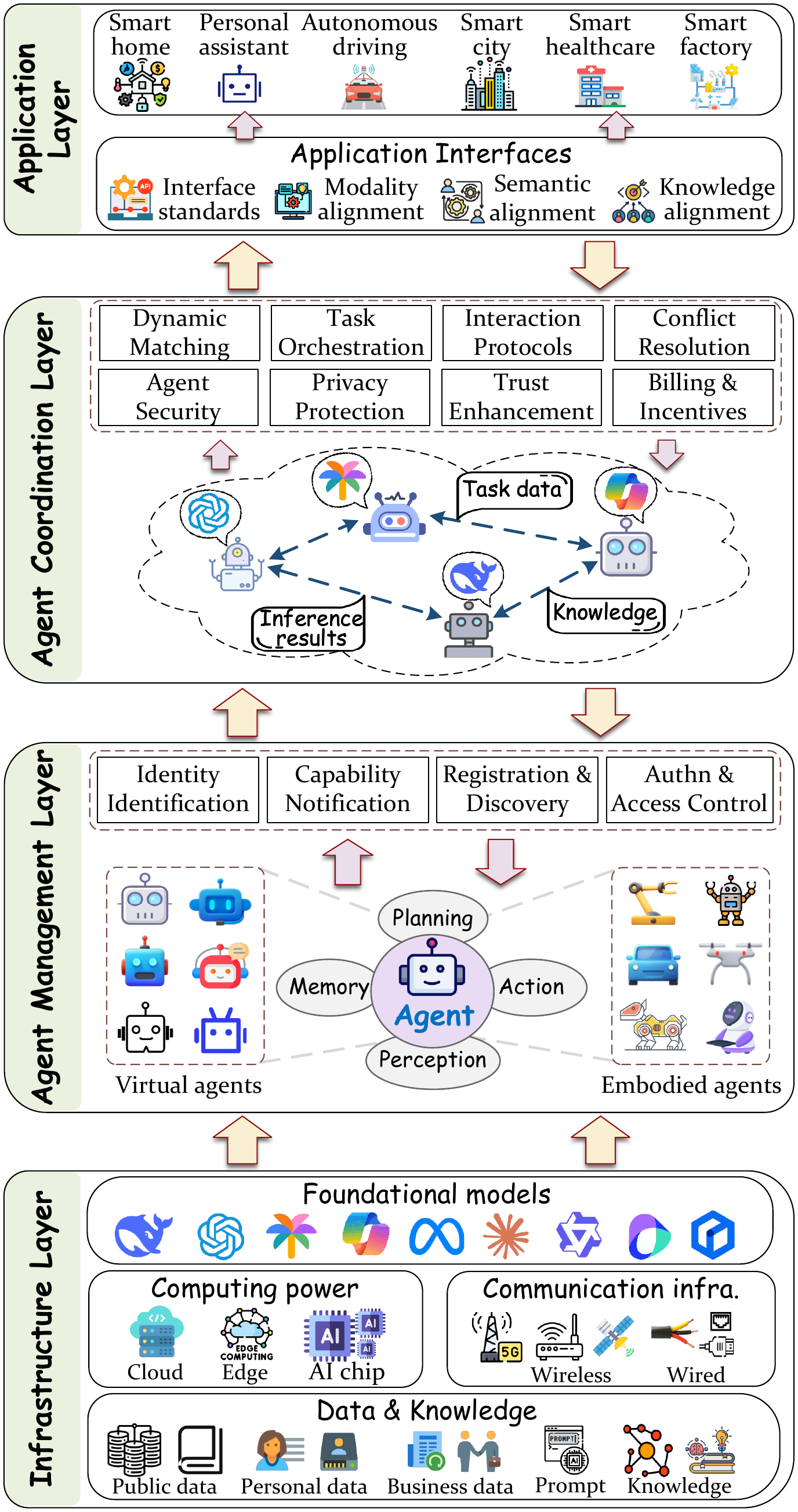}
 \caption{General architecture of IoA, comprising four tiers from bottom to top: the infrastructure layer, agent management layer, agent coordination layer, and agentic application layer.}\label{fig:archi}\vspace{-3.5mm}
\end{figure}

\textit{4) General IoA Architecture.} As shown in Fig.~\ref{fig:archi}, the hierarchical IoA architecture comprises the following four layers that enable scalable, secure, and interoperable agent collaboration. 

\begin{itemize}
    \item \textit{The infrastructure layer} serves as the foundation of the IoA, integrating critical resources across AI models, data, knowledge, computing, and communication \cite{chen2025ioa}. Foundational models such as DeepSeek-R1, GPT-4o, and LLaMA 4 function as the cognitive core of agents operating on heterogeneous platforms including cloud-based GPU/TPU clusters, edge nodes, and terminal devices such as UAVs and robots. 
    The data infrastructure supports a wide range of multimodal inputs, including text, sensor streams, 3D point clouds, and real-time flows, structured as public or private assets and enriched by prompt repositories and both domain-specific and proprietary knowledge bases. Reliable, low-latency connectivity is enabled by various communication technologies, including 5G URLLC, mesh networks for autonomous agents, and satellite-based backup links.
    
    \item \textit{The agent management layer} manages the identity, capabilities, discovery, and lifecycle of agents within the IoA. Identity mechanisms should support autonomous identification and secure cross-domain authentication, such as decentralized identifiers (DIDs) \cite{halpin2020vision}. Agent capabilities are multi-dimensional (encompassing functional, contextual, and domain-specific attributes) and can be described using semantic models that support dynamic updates and machine-readable discovery. Besides, flexible communication frameworks, such as publish-subscribe (pub/sub) mechanisms \cite{hong2024metagpt}, facilitate capability announcements, while distributed registries and context-aware discovery mechanisms \cite{bhopale2024transformer, zhang2025toward} (e.g., intent-based matching) enable intelligent agent registration and task-oriented agent matching.

    \item \textit{The agent coordination layer} manages distributed task execution and agent collaboration. It supports task decomposition \cite{wei2023cot,yao2023tot,besta2024graph}, flow orchestration \cite{liang2024encouraging}, adaptive communication protocols \cite{Agora}, and dynamic agent-task matching \cite{wu2024autogen}. Furthermore, conflict resolution and consensus mechanisms \cite{amirkhani2022consensus}, such as Byzantine fault tolerance and arbitration, ensure robustness in adversarial environments. Additional functions include trust management, billing, security safeguards, and privacy-preserving mechanisms, which collectively provide a secure, reliable, and accountable foundation for large-scale multi-agent coordination.

    \item \textit{The agent application layer} offers standardized interfaces and semantic alignment to support interoperable, domain-specific agent ecosystems \cite{aminiranjbar2024dawn}. It facilitates cross-modal, semantic, and knowledge-level integration, allowing agents to function seamlessly across diverse and heterogeneous environments. Representative IoA applications include traffic optimization and emergency response in smart cities \cite{wu2025multi}, predictive maintenance and supply chain management in smart factories \cite{10090432}, and collaborative surgical robotics and diagnostics in healthcare \cite{qiu2024llm}.
\end{itemize}

\subsection{Key Characteristics of IoA}\label{subsec:Characteristic} 
Generally, IoA exhibits the following unique characteristics. 

\textit{1) Autonomous Intelligence.} In the IoA paradigm, agents operate with a high degree of autonomy, capable of proactively advertising their capabilities and independently initiating collaborations  \cite{chen2025ioa,A2A}. They can dynamically orchestrate task flows, negotiate interaction protocols, resolve conflicts, and reach consensus without human intervention. In contrast, traditional Internet systems rely on rigid, pre-defined protocols and APIs tailored to specific applications, while IoT devices offer limited automation and intelligence. 
IoA agents, powered by large models, transcend these limitations by enabling autonomous capability discovery, adaptive reasoning, and real-time task orchestration. Key mechanisms include:
\begin{itemize}
    \item \textit{Proactive Capability Discovery:} Agents autonomously register at gateways by advertising capacities \cite{aminiranjbar2024dawn} and locating complementary partners for task collaboration on demand.
    \item \textit{Autonomous Workflow Orchestration \& Interaction Topology Optimization:} By continuously negotiating and re-planning, agents autonomously decompose tasks \cite{wei2023cot}, assign subtasks \cite{wu2024autogen}, and adjust execution flows \cite{liang2024encouraging}. Besides, they autonomously optimize interaction topologies \cite{liu2024dynamic}, e.g., deciding agent turn-taking in negotiations, to enhance efficiency in dynamic environments.
    \item \textit{Adaptive Conflict Resolution and Consensus:} Through contextual negotiation, agents can autonomously detect and resolve conflicts, reconcile competing goals, and reach agreement on consensus strategies \cite{chen2025ioa,holt2023l2mac}.
\end{itemize}

\textit{2) High Dynamics.} IoA networks are inherently dynamic, characterized by the on-demand creation, migration, and termination of virtual agents, as well as the physical mobility of embodied agents such as UAVs and robots. This agility enables real-time reconfiguration of agent teams to accommodate varying task demands, environmental conditions, or resource constraints. 
Additionally, agent's capabilities, as well as task orchestration and interaction protocols, adjust continuously. 
\begin{itemize}
    \item \textit{On-demand Agent Lifecycle:} Agents dynamically spin up or retire based on task load or resource availability. For instance, a customer-service agent cluster expands during peak sales events and contracts once demand subsides.
    \item \textit{Evolving Agent Capability:} Agent capabilities are also context-dependent, adapting based on current task goals, tool availability, and its functional profile (e.g., sensor suite and reasoning engine). For instance, a ground robot might deploy its camera and LiDAR fusion pipelines to navigate complex terrain, yet switch to natural‐language processing (NLP) modules when interpreting spoken commands, exhibiting differentiated capability.
    \item \textit{Real-Time Workflow Reconfiguration:} The task workflow orchestration and interaction strategies among agents are highly dynamic, requiring agents to continually adapt their roles, behaviors, and communication patterns in real time to maintain effective collaboration \cite{tran2025multi}. For instance, an autonomous delivery fleet encountering traffic congestion can renegotiate routes, reassign deliveries, and update messaging patterns instantly.
\end{itemize}

\textit{3) High Heterogeneity.} IoA encompasses a wide spectrum of virtual and embodied agents, ranging from microcontroller nodes and autonomous UAVs to GPU cluster-driven agents and home robots, each with varying model architectures and distinct compute, energy, sensing, and communication capabilities. These agents operate over heterogeneous networks (e.g., cellular, satellite, and mesh Wi-Fi) with varying access modes and link types \cite{wang2024large}, forming a heterogeneous ecosystem that challenges seamless interoperability. 
\begin{itemize}
    \item \textit{Capability Variance:} Agents range from battery-constrained end devices to large model agents with abundant GPU and memory. Task assignments and workload offloading should dynamically adapt to agents' resource constraints.
    \item \textit{Perceptual Modality Diversity:} Data formats span LiDAR point clouds, hyperspectral imagery, biosensor streams, and language embeddings. Seamless collaboration requires on-the-fly modality conversion and alignment, such as converting point-cloud scans into semantic graphs for planning agents.
    \item \textit{Communication Heterogeneity:} Agents communicate over multiple communication links. Interoperability demands multi-layer protocol stacks and runtime adapters to negotiate bandwidth, latency, and security trade-offs across diverse communication types.
\end{itemize}

\textit{4) Large-scale Scalability.}
IoA should scale from small ad-hoc agent teams to billions of agents across distributed domains. This demands elastic IoA architectures that support real-time discovery, grouping, and reconfiguration for vast agent populations under dynamic workloads.
\begin{itemize}
    \item \textit{Hierarchical Sharding:} Agents self-organize into hierarchical clusters or ``shards'' based on geography, function, or task affinity \cite{li2024survey}. For instance, domain gateways host proxy agents that register local agents, broadcast their capabilities, and coordinate cross-domain collaboration and state synchronization via relay gateways, as shown in Fig.~\ref{fig:overview}.
    \item \textit{Elastic Resource Orchestration:} A multi-tier orchestration layer dynamically allocates compute, storage, and communication resources across cloud, edge, and device nodes. Agents experiencing bursts of demand can seamlessly migrate workloads to underutilized peers or cloud/edge agents without service interruption.
    \item \textit{Adaptive Overlay Networks:} IoA dynamically constructs task-oriented overlay networks \cite{liu2024dynamic}, such as transient micro-swarms for disaster mapping or specialized maintenance teams in smart factories, that self-assemble when a mission begins and dissolve once it ends.
    By tailoring peer-to-peer (P2P) connections to the specific task, these overlays optimize routing efficiency, reduce unnecessary traffic, and limit global network churn \cite{ANP}.
    \item \textit{Distributed Fault Tolerance \& Self-Healing:} At large scales, agent failures are inevitable. IoA integrates lightweight protocols to continuously monitor peer health. When an outage is detected, agents automatically reroute tasks around offline nodes and redistribute workloads, ensuring the system adapts in real time and maintains overall operational integrity.
\end{itemize}

\textit{5) Semantic-aware Communication.} Powered by large models, agents in IoA possess advanced context-aware semantic understanding and reasoning capabilities, enabling them to interpret intent, adapt communication strategies, and align meanings in real time. Unlike traditional Internet or IoT systems, which prioritize reliable data transmission, agent-based communications are inherently task-driven and semantically enriched, which exhibits three distinctive characteristics \cite{10798108}:
\begin{itemize}
    \item \textit{Computing-Oriented Communication:} Rather than transmitting raw data, agents prioritize exchanging context-aware and task-relevant information to support intelligent coordination. Agent communications achieve this by first synchronizing knowledge base through the network before communication. For instance, an agent may align with a shared knowledge base to interpret sensor inputs and transmit only essential inferences, thereby reducing communication overhead.
    \item \textit{Persistent Communication:} Agent collaboration often span extended durations, requiring continuous adaptation to evolving task states. Agent communications leverage persistent memory to maintain shared context (e.g., incremental knowledge updates), enabling agents to reference prior interactions efficiently.
    \item \textit{Memory-Based Communication:} Agent tasks often exhibit sequential reasoning or iterative refinement \cite{yao2023ReAct,shinn2023Reflexion}. Agents leverage memory to compress and contextualize messages, such as transmitting only updates relative to previously shared knowledge.
\end{itemize}

This shift from data-centric to meaning-aware communication is fundamental to building goal-oriented, adaptive, and scalable MAS.

\begin{table*}[!t]
\centering\setlength{\abovecaptionskip}{0cm}
\caption{Key Comparisons between Traditional Internet, IoT, and IoA}
\label{tab:internet_vs_ioa}
\begin{tabular}{lccc}
\toprule
\textbf{}                         & \textbf{Traditional Internet} & \textbf{Internet of Things (IoT)} & \textbf{Internet of Agents (IoA)}     \\ \midrule
\textbf{Core Objective}         & Host \& Information Connectivity & Device \& Information Connectivity & Agent \& Knowledge Connectivity          \\ \hline
\textbf{Service Objects}          & Humans                        & Smart Devices (Sensors/Actuators) & Autonomous Agents (Virtual/Embodied)  \\ \hline
\textbf{Architecture}           & Centralized (Client-Server)      & Decentralized (End-Edge-Cloud)       & Hybrid (P2P + Proxy-based)      \\ \hline
\textbf{Addressing Mechanism}     & IP                            & IP, Static Device Identity        & IP, Dynamic Semantic Identifiers      \\ \hline
\textbf{Interaction Mode}       & Passive (Request-Response)       & Event-Driven (Trigger-Based)       & Proactive (Goal/Task-Oriented)           \\ \hline
\textbf{Communication Level}      & Bit-Level Transmission        & Bit-Level + Lightweight Protocols & Semantic-Level Exchange               \\ \hline
\textbf{Autonomy Source}          & Human-Controlled              & Rule-based Device Logic           & Large Model-Driven Agent Intelligence \\ \hline
\textbf{Network Dynamics} & Low (Static Topologies)          & Medium (Dynamic Topologies)           & High (Evolving Interactions \& Mobility) \\ \hline
\textbf{Standardization Maturity} & Well-Established              & Stabilized with Evolution         & Emerging \& Evolving                  \\ \bottomrule
\end{tabular}
\end{table*}

\begin{table*}[!t]
\centering\setlength{\abovecaptionskip}{0cm}
\caption{{Conceptual Comparison of IoA with Related Paradigms}}\label{tab:IoA_vs_others}
\begin{tabular}{p{1.6cm}|p{3.5cm}|p{2.2cm}|p{3.4cm}|p{3.3cm}}
\toprule
\textbf{Paradigm} & \textbf{Scope \& Focus} & \textbf{Scale} & \textbf{Autonomy \& Adaptivity} & \textbf{Typical Applications} \\ \hline
\textbf{MAS} & Coordinated problem-solving in bounded environments & Moderate, domain-specific & Agents collaborate under centralized/semi-centralized orchestration & \makecell[l]{Home robots coordination, \\software development} \\ \hline
\textbf{Agentic Web} & Intelligent services on top of the Web & Large, web-centric & Enhancing user services via intelligent agents & \makecell[l]{Personalized assistants, \\intelligent web navigation}  \\ \hline
\textbf{Web 3.0} & Decentralized, semantic, and ownership-driven web & Internet-wide & Limited autonomy; emphasis on data interoperability & \makecell[l]{Decentralized finance, \\semantic knowledge sharing}  \\ \hline
\textbf{IoA} & Internet-like fabric interconnecting heterogeneous agents & Large-scale, cross-domain & High autonomy, dynamic adaptivity, open collaboration & Cross-domain agent ecosystems \\ \bottomrule
\end{tabular}
\end{table*}

\textit{6) Task-driven Cooperation.} IoA prioritizes task-oriented networking, where agents dynamically align their capabilities to task requirements and form ad hoc teams for autonomous cross‐domain operations. 
\begin{itemize}
    \item \textit{Agent-Task Matching:} Tasks are published to a decentralized registry enriched with semantic metadata \cite{aminiranjbar2024dawn}. Agents autonomously announce their capabilities and are matched in real time based on criteria such as current workload, geographic proximity, reliability, and quality-of-service (QoS) metrics.
    \item \textit{Task-oriented Team Formation:} Once matched, agents negotiate roles, priorities, and execution flows to assemble an optimal team \cite{li2023camel}. Teams then instantiate ephemeral overlay networks such as micro-swarms that coordinate subtasks via autonomous negotiation. 
\end{itemize}

\textit{Summary:} The autonomous intelligence of agents in IoA enables dynamic team formation, mirroring how human teams are assembled on demand. Agents can autonomously negotiate communication protocols, leverage diverse tools to facilitate interaction, and resolve conflicts without human intervention. However, this autonomy, combined with the inherent heterogeneity and high dynamics of the IoA ecosystem, introduces substantial challenges in agent discovery, capability advertisement, interoperability, and the design of effective economic incentive mechanisms. Moreover, the task-driven nature of agents and their reliance on semantic communication necessitate context-aware coordination and intelligent interpretation of intent, further complicating system design. 
This shift from rule-based automation and data-centric communication to large model-driven intelligence and meaning-aware communication is fundamental toward self-organizing and goal-oriented IoA ecosystems.

\subsection{Key Differences Among Internet, IoT, and IoA}\label{subsec:Differences}
IoA represents a paradigm shift from the traditional Internet and IoT by fundamentally redefining core objectives, architectural principles, operational mechanisms, and interaction patterns, as summarized in Table~\ref{tab:internet_vs_ioa}.

\textit{1) Core objectives} of the Internet, IoT, and IoA:
\begin{itemize}
    \item \textit{Traditional Internet} primarily focuses on global host and information connectivity via the TCP/IP protocol and the World Wide Web (WWW). It supports essential services such as web browsing, file transfer, and email, prioritizing reliable information delivery and human-centric experiences. 
    
    \item \textit{IoT} focuses on device monitoring and control, interconnecting sensors and actuators with analytics platforms to automate data collection and feedback loops.
    
    \item \textit{IoA} emphasizes autonomous agent collaboration and knowledge exchange, aiming to form dynamic self-organizing networks driven by tasks \cite{chen2025ioa}. Agent interactions are inherently fault-tolerant, with a focus on minimizing latency. The primary goal is to connect agents with one another, enabling capability notification, dynamic matching, and task orchestration.
\end{itemize}

\textit{2) Architectures} of the Internet, IoT, and IoA:
\begin{itemize}
    \item \textit{Traditional Internet} relies on centralized servers, e.g., clouds and domain name system (DNS) servers, and follows a client-server architecture, where users access resources via fixed IP addresses or domain names, and interactions are typically passive and request-driven.

    \item \textit{IoT} uses a decentralized end-edge-cloud architecture, often with brokered or P2P messaging (e.g., MQTT, CoAP) \cite{al2015internet,thangavel2014performance}, allowing devices to offload processing to edge or cloud nodes.  

    \item \textit{IoA} adopts a hybrid decentralized architecture \cite{aminiranjbar2024dawn}, combining P2P overlays and proxy gateway agents, enabling agents to join, leave, and collaborate across domains on demand. Agents can dynamically join or leave the network. Communication is driven by semantic interaction protocols and task-oriented networking, enabling agents to announce capabilities and initiate collaborations. 
\end{itemize}

\textit{3) Addressing \& discovery mechanisms} among the Internet, IoT, and IoA:
\begin{itemize}
    \item \textit{Traditional Internet} relies on static IP addresses and DNS resolution, binding identifiers to physical locations or service entry points (e.g., 192.168.1.1 or www.example.com). Resource discovery is typically via manual uniform resource locator (URL) lookup or search engines. 

    \item \textit{IoT} extends IP addressing with static device identifier and supports automated discovery using protocols such as multicast DNS (mDNS) and CoAP resource directories, and lightweight service discovery \cite{al2015internet,thangavel2014performance}.  

    \item \textit{IoA} supports dynamic dynamic agent identifiers and capability labels, via methods such as semantic identifiers \cite{fernandez2021semantic} and semantic index \cite{kiryakov2004semantic}, that update in real time according to agent states or task requirements. Discovery is guided by capability modeling languages, enabling proactive announcements (e.g., via pub/sub) and on-demand and task-driven matching through context-aware retrieval.
\end{itemize}

\textit{4) Interaction modes} among the Internet, IoT, and IoA:
\begin{itemize}
    \item Interactions on the \textit{traditional Internet} are primarily human-driven and built on passive request-response exchanges via GUIs and predefined service interfaces (e.g., APIs), with human endpoints initiating most transactions.

      \item Interactions on the \textit{IoT} are event-driven and trigger-based machine-to-machine messaging, enabling basic automation but limited by static schemas and minimal semantics.  
 
    \item \textit{IoA} enables proactive and goal-oriented interactions, often using natural language or structured semantic messages. Agents can negotiate protocols, adapt behavior, and coordinate autonomously without human intervention. 
\end{itemize}

{While the IoA is inspired by earlier paradigms, it is distinct in scope and objectives. Traditional MAS primarily focuses on coordinated problem-solving within bounded environments, often under centralized or semi-centralized orchestration. In contrast, the IoA emphasizes large-scale, open, and heterogeneous ecosystems where autonomous agents interact dynamically across domains, infrastructures, and ownership boundaries. Similarly, the Agentic Web highlights intelligent services built on top of existing web platforms, and Web 3.0 mainly advances decentralized data ownership and semantic interoperability. The IoA builds upon but goes beyond these paradigms by integrating autonomy, adaptivity, and economic incentives into a unified, Internet-like fabric for agent collaboration. Table~\ref{tab:IoA_vs_others} summarizes key conceptual differences of IoA with MAS, Agentic Web, and Web 3.0.}

\subsection{Emerging IoA Application Scenarios}\label{subsec:Prototypes}

\textit{1) Typical Multi-Agent Frameworks:} Recent advancements in LLMs have spurred the development of various agent frameworks capable of autonomous or semi-autonomous task execution through advanced reasoning, planning, and tool integration. 
AutoGPT \cite{AutoGPT} implements autonomous goal-driven agents by iteratively refining actions through recursive self-prompting. 
AutoGen \cite{wu2024autogen} introduces conversational agent networks where multiple LLM-powered agents (e.g., assistants and user proxies) interact dynamically, supporting both human-in-the-loop and fully automated workflows. 
LangChain \cite{LangChain} provides a modular framework for chaining LLM calls with external tools (e.g., APIs and databases) and memory systems, offering flexibility for customizable agent pipelines. 
MetaGPT \cite{hong2024metagpt} simulates software development teams with role-based agents (e.g., product managers and engineers) that follow structured workflows, enhancing multi-agent collaboration. 
Other notable frameworks include BabyAGI \cite{BabyAGI}, which focuses on task-driven autonomous iteration, and CAMEL \cite{li2023camel}, which explores role-playing agent societies with communicative agents.  
Table~\ref{tab:agent_frameworks} shows the comparisons of mainstream open-source multi-agent frameworks. 

\begin{table*}[!t]
\centering\setlength{\abovecaptionskip}{0cm}
\caption{Comparison of Mainstream Agent Frameworks}
\label{tab:agent_frameworks}
\begin{tabular}{lllll}
\toprule
\textbf{Framework} & \textbf{Key Features} & \textbf{Strengths} & \textbf{Weaknesses} & \textbf{Primary Use Case} \\
\midrule
MetaGPT\cite{hong2024metagpt} & 
\makecell[l]{Role-based agents,\\ structured workflows,\\ multi-agent collaboration} & 
\makecell[l]{High efficiency in\\ complex tasks} & 
\makecell[l]{Require predefined roles, \\less flexible} & 
\makecell[l]{Automated software\\ development} \\
\midrule
LangChain\cite{LangChain} & 
\makecell[l]{Modular LLM chaining,\\ memory, tool integration} & 
\makecell[l]{Highly customizable,\\ supports diverse tools} & 
\makecell[l]{Steeper learning curve,\\ manual tuning needed} & 
\makecell[l]{Custom agent pipelines,\\ RAG systems} \\ \midrule
AutoGPT\cite{AutoGPT} & 
\makecell[l]{Recursive self-prompting, \\goal-driven autonomy} & \makecell[l]{Fully autonomous \\task execution} & \makecell[l]{Prone to loops, \\high compute cost} & \makecell[l]{General task automation} \\
\hline
AutoGen\cite{wu2024autogen} & \makecell[l]{Conversational MAS, \\human-in-the-loop support} & \makecell[l]{Flexible collaboration, \\dynamic interactions} & \makecell[l]{Complex setup for \\optimal performance} & \makecell[l]{Multi-agent dialogue, \\AI teamwork} \\
\hline
BabyAGI\cite{BabyAGI} & \makecell[l]{Task-driven autonomous \\iteration, simple architecture} & \makecell[l]{Lightweight, \\easy to deploy} & Limited reasoning depth & Small-scale automation \\
\hline
CAMEL\cite{li2023camel} & \makecell[l]{Role-playing agent societies, \\communicative agents} & \makecell[l]{Simulates human-like \\interactions} & \makecell[l]{Less optimized for tool use} & \makecell[l]{Social AI, \\research simulations} \\
\bottomrule
\end{tabular}
\end{table*}

\begin{figure}[!t]
\centering \setlength{\abovecaptionskip}{-0.cm}
\includegraphics[width= 1.03\linewidth]{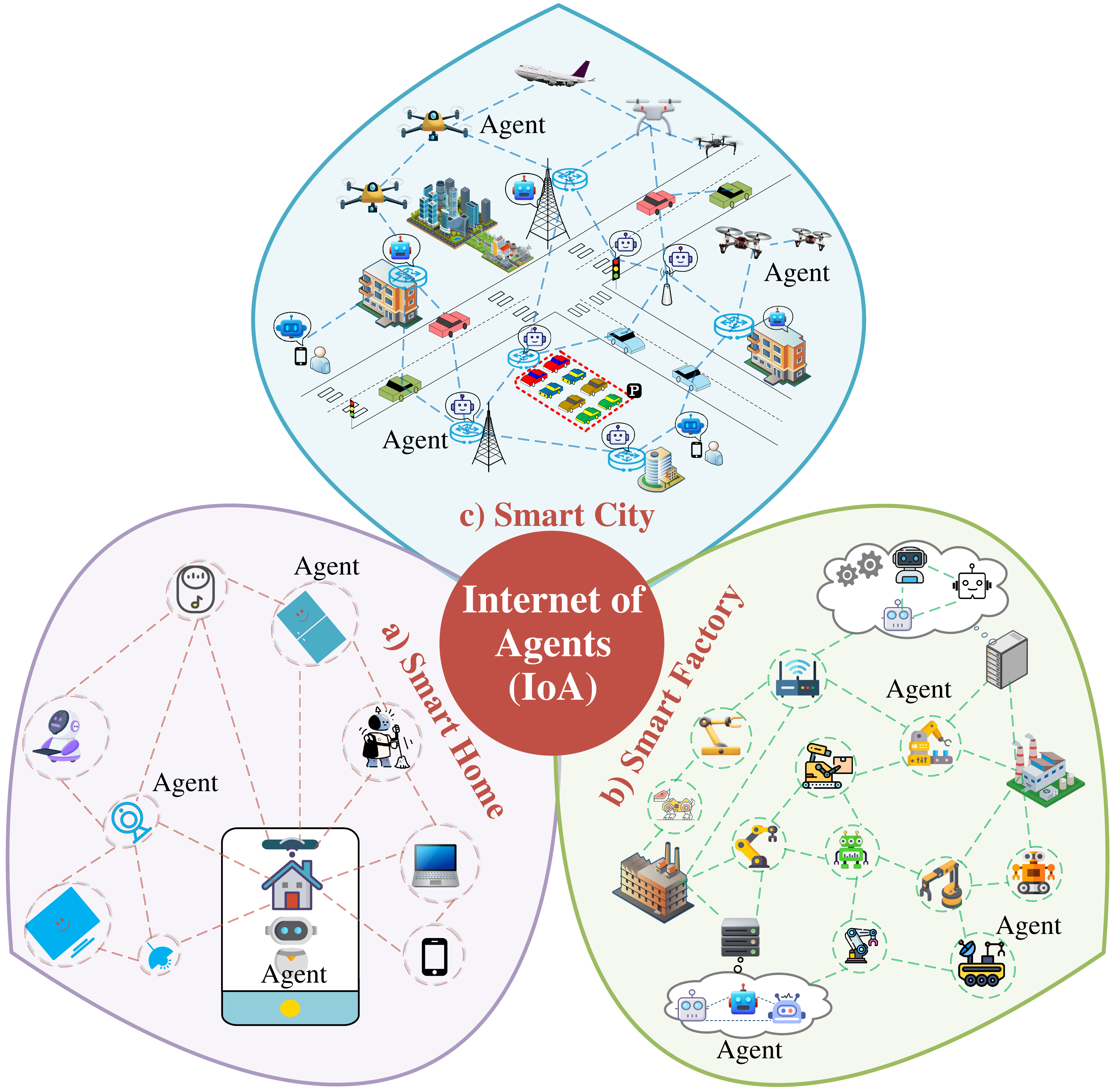}
 \caption{An overview of IoA applications: (a) smart city, (b) smart home, and (c) smart factory.}\label{fig:application}\vspace{-3mm}
\end{figure}

\textit{2) Roadmap:} 
Existing representative industrial practices include computer use, AI smartphones, Anthropic's model context protocol (MCP) \cite{MCP}, and Google's agent-to-agent (A2A) Protocol \cite{A2A}. (i) Computer-Use Agents (CUAs): As existing Internet services are originally built for humans, CUAs \cite{sager2025ai} such as OpenAI's Operator mimic human GUI interactions (e.g., mouse clicks and keystrokes) to drive browsers and mobile interfaces, but incur significant screen-scraping overhead that undermines agent efficiency. (ii) AI Smartphones: AI smartphones (e.g., Apple Intelligence) enable cross-App data access and function calls by exposing App APIs to agents, better aligned with agent's strength in processing raw data than GUI emulation. (iii) MCP places large models at the core of the ecosystem, treating the entire Internet as its contextual knowledge base and toolkit. Standardized APIs, dynamic plugin support, and persistent context management allow a single model to orchestrate across diverse services. (iv) A2A protocol aims to form a decentralized agent network where agents autonomously discover, negotiate, and collaborate with each other.
Ultimately, IoA should provide (1) seamless interoperability to eliminate data silos, (2) agent-native interfaces for direct efficient inter-agent interactions, and (3) protocol-driven self-organizing cooperation among agents. 

{\textit{3) Real IoA Prototypes and Industrial Case Studies:} To illustrate the practical realization of the IoA, we analyze representative academic and industrial IoA implementations.}

{\textit{a) IoA Prototype in \cite{chen2025ioa}.} Chen \textit{et al.} present an IoA prototype to connect heterogeneous virtual agents for collaborative intelligence. It adopts a layered client-server architecture with interaction, data, foundation layers, comprising four core components: an agent registry \& discovery module for registration and semantic matching; a task management \& orchestration engine for autonomous team formation, subtask allocation, and workflow sequencing; a data/knowledge store that enables context-aware information sharing and local agent memory; and message-protocols that specify agent message formats, group-chat semantics, and a finite-state conversation controller inspired by speech-act theory. By integrating symbolic, LLM-powered, and task-specialized agents, this prototype supports dynamic teaming and conversational flow control rather than fixed, hard-coded pipelines.
}

{\textit{b) DAWN \cite{aminiranjbar2024dawn}.} DAWN aims to support globally distributed IoA collaboration via a hierarchical architecture comprising \textit{Principal Services Agents} and \textit{Gateway Services Agents}. The Principal Agent plans and orchestrates tasks, while each associated Gateway Agent maintains registries of tools, agents, and applications, exposing them via RESTful endpoints. Upon receiving a task, the Principal Agent delegates subtasks to Gateway Agents, which identify suitable resources and return them for execution.
DAWN supports three operational modes: \textit{No-LLM mode} for highly predictable, deterministic tasks using traditional tools and human-designed workflows; \textit{Copilot mode}, where agents assist humans with LLM-guided planning and resource orchestration; and \textit{LLM agent mode}, in which agents operate autonomously, performing reasoning, planning, and execution using LLMs with minimal human intervention.}

{\textit{c) Siemens Industrial Copilot \cite{SiemensAgent}.} Unveiled at 2025 in Detroit, Siemens’ Industrial Copilot framework applies autonomous agents across the entire production lifecycle, supporting design optimization, engineering assistance, and predictive maintenance in smart factories. In the design phase, design agents assist engineers with system modeling and workflow optimization. During engineering, planning and engineering agents automate routine programming, system configuration, and cross-team coordination, reducing manual effort and errors. Finally, in maintenance, operation and service agents analyze real-time data to anticipate equipment failures and enable proactive scheduling, supporting predictive maintenance strategies. This multi-agent copilot ecosystem collectively enhances efficiency, reliability, and resilience in industrial automation.}

{\textit{d) Midea Intelligent-Agent Factory \cite{MideaAgent}.} Midea's Jingzhou facility exemplifies an experimental intelligent-agent factory powered by the Factory Brain: a central orchestrator that enables A2A-based multi-agent coordination and leverages an industrial-grade large-model inference engine. The system deploys 14 agents across 38 production scenarios, spanning scheduling, energy management, quality assurance, operations, and maintenance. For instance, in automated quality inspection, quality and R\&D agents collaborate via the Factory Brain to implement closed-loop inspections using AI-enabled glasses, reducing first-pass inspection time from 15 minutes to 30 seconds. 
Embodied humanoid agents such as Meiro are also used to autonomously execute inspections, equipment checks, and safety patrols. Together, these heterogeneous agents demonstrate the potential of coordinated IoA ecosystems to enhance efficiency and flexibility in manufacturing.
}

\textit{4) Applications:} The IoA paradigm holds transformative potential across diverse sectors. 
As shown in Fig.~\ref{fig:application}, we illustrate its applications in five representative scenarios: smart homes, healthcare, smart grids, smart factories, and smart cities.
\begin{itemize}
    \item \textit{Scenario 1: Agent Communications within an IoA Subnet in Smart Home.} In the smart home context, a dedicated IoA subnet enables diverse domestic agents, such as housekeeping robots, digital life assistants, robotic pets, and smart appliances, to automatically discover one another and establish task-specific P2P overlays. Upon joining the home network, a newly activated housekeeping robot obtains a unique digital identity, identifies co-located agents, and dynamically forms task groups to coordinate functions such as environmental monitoring, meal preparation, and energy management. Agents autonomously migrate between Wi-Fi and cellular access on-the-fly while leveraging IoA subnet services for multi-modal data forwarding and intra-domain resource sharing, thereby delivering highly adaptive, self-orchestrating living environment. 
    
        \item {\textit{Scenario 2: Healthcare Robot Coordination within an IoA Subnet in Medical Facilities.} In the healthcare context, a dedicated IoA subnet interconnects diverse medical agents, such as surgical robots, anesthesia robots, monitoring robots, and delivery robots, to enable seamless intra-facility coordination. Upon detecting abnormal patient vital signs, a monitoring robot leverages the subnet to discover surgical robots for instrument preparation, anesthesia robots for dosage calculation, and delivery robots for timely medication transport. Each participating agent is assigned a unique digital identity and engages in encrypted, credential-based communication to ensure data integrity and patient confidentiality. By dynamically forming task-specific teams and orchestrating multi-agent workflows, the subnet enhances emergency responsiveness, reduces procedural delays, and improves overall patient safety within medical facilities.}
    
    \begin{figure*}[!t]
\centering \setlength{\abovecaptionskip}{-0.cm}
\includegraphics[width= 0.7\linewidth]{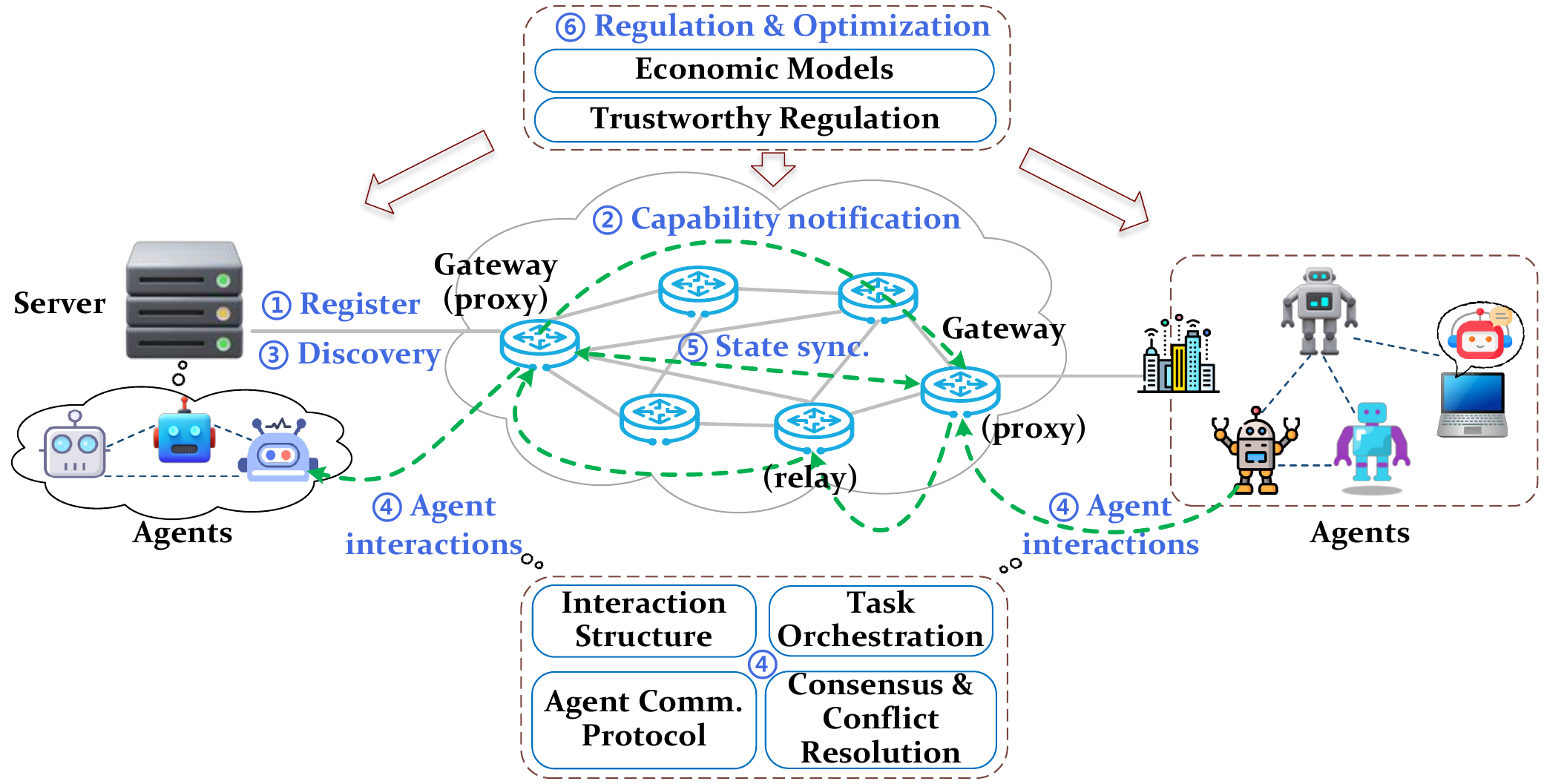}
 \caption{An overview of cross-domain agent interaction lifecycle in IoA, consisting of six phases: \ding{172}agent register; \ding{173}capability notification (in Sect.~\ref{subsec:Paradigm1}); \ding{174}agent discovery (in Sect.~\ref{subsec:Paradigm1}); \ding{175}agent interactions with the help of proxy gateways and relay gateways, including interaction structure design (in Sect.~\ref{subsec:Paradigm2}), task orchestration (in Sect.~\ref{subsec:Paradigm2}), agent communication protocol design (in Sect.~\ref{subsec:Paradigm3}), and consensus \& conflict resolution (in Sect.~\ref{subsec:Paradigm4}); \ding{176}state synchronization; and \ding{177} regulation \& optimization, including economic models (in Sect.~\ref{subsec:Paradigm5}) and trustworthy regulation (in Sect.~\ref{subsec:Paradigm6}).}\label{fig:workflowexample}\vspace{-2mm}
\end{figure*}

        \item \textit{Scenario 3: Coordination between Agents in an IoA Subnet and and External Agents in Smart Factory.} Within a smart factory, the IoA subnet connects on-site manufacturing agents (e.g., robotic arms, autonomous guided vehicles, and digital twin controllers) with external ecosystem participants (such as component suppliers, logistics UAVs, and cloud analytic agents). For instance, an online production supervisor agent can request spare parts through a supplier's digital assistant, assign an autonomous forklift to negotiate pickup timing with a delivery UAV, and authenticate identities via QR-code or device-to-device (D2D) communications. Throughout this process, the network provides digital identity verification and sensing-based path-planning services, ensuring that materials arrive just-in-time and production workflows adapt responsively to supply-chain dynamics. 

        \item {
        \textit{Scenario 4: Energy Management via IoA Subnet and External Agents in Smart Grid.}
Within smart grid ecosystems, an IoA subnet interconnects domain-specific energy agents, such as renewable generator agents, storage system agents, smart meter agents, and demand controller agents. These agents coordinate locally to balance electricity production, storage, and consumption in real time. At the same time, the subnet interacts with external entities including regional grid operators, energy trading platforms, and forecasting services. For instance, renewable generator agents negotiate surplus electricity allocation with external market agents, while storage agents synchronize with operator agents to provide ancillary services such as peak shaving and frequency regulation. Meanwhile, smart meter and demand controller agents exchange data with forecasting services to anticipate consumption fluctuations and dynamically adjust appliance usage. Through this integration, the IoA subnet ensures resilient, demand-responsive energy distribution, while external collaborations enhance system-wide efficiency and economic optimization.}
        
    \item \textit{Scenario 5: Cross-Domain Agent Networking in Smart City.} At the urban scale, IoA fosters dynamic cross-domain collaboration among heterogeneous agents from diverse stakeholders, such as municipal traffic controllers, public safety UAVs, autonomous vehicles, and emergency responder robots. By forming on-demand task-specific teams secured via mutual identity authentication, agents from different entities can rapidly assemble to conduct real-time surveillance, detect anomalies (e.g., fires or traffic incidents), and coordinate multimodal responses. Upon task completion, resources are released automatically, optimizing network load and ensuring scalable and resilient orchestration of city-wide services. 
\end{itemize}

{\textit{Summary and Lessons Learned:} This subsection first surveys existing multi-agent frameworks, ranging from goal-driven systems (e.g., AutoGPT and BabyAGI) to modular pipelines (e.g., LangChain) and role-based simulations (e.g., MetaGPT), highlighting advances in inter-agent reasoning, planning, and tool integration. We then sketch a technology roadmap, ranging from GUI-driven CUAs and AI smartphones to MCP and A2A protocols and future IoA protocols. Finally, we illustrate several representative IoA application scenarios: P2P agent communication (in smart homes and smart healthcare), coordination between on-site and external agents (in smart factories and smart grids), and cross-domain networking (in smart cities). Collectively, these developments underscore the need for seamless interoperability, agent-native interfaces, and protocol-driven self-organization as the foundation for future IoA. 
}

\section{Working Paradigms of Internet of Agents}\label{sec:Paradigms}
In this section, we investigate the enabling mechanisms of IoA under the layered IoA architecture, including capability notification \& discovery (in Sect.~\ref{subsec:Paradigm1}), task orchestration \& assignment (in Sect.~\ref{subsec:Paradigm2}), agent communication protocol (in Sect.~\ref{subsec:Paradigm3}), consensus \& conflict resolution (in Sect.~\ref{subsec:Paradigm4})), economic models (in Sect.~\ref{subsec:Paradigm5}), and trustworthy regulation (in Sect.~\ref{subsec:Paradigm6}). 
Fig.~\ref{fig:workflowexample} illustrates the detailed cross-domain agent interaction workflow in IoA. 

\subsection{Capability Notification \& Discovery in IoA}\label{subsec:Paradigm1}

\subsubsection{Capability Evaluation of Agents}
Accurately evaluating the capabilities of agents is foremost for effective task assignment and dynamic collaboration in IoA. Unlike traditional MAS, large model-based agents exhibit highly dynamic and evolving abilities, such as diverse reasoning paradigms and adaptive tool utilization. 
\begin{itemize}
    \item \textit{Self-reported Capability Declaration.} In IoA, each newcomer agent provides a self-reported capability profile during registration, detailing its foundational model characteristics, available tools, and functional expertise \cite{chen2025ioa,A2A}. For instance, an agent might declare its profile as ``\{model: ``GPT-4o'', tools: [``Web Browser'', ``Code Executor''], description: ``Specialized in QA tasks''\}''. While self-reporting facilitates rapid onboarding, it is inherently prone to inaccuracies or overstatements. 
    \item \textit{Systematic Capability Verification.} To enhance reliability, gateway agents could perform capability verification using standardized evaluation suites to objectively assess critical competencies. For instance, reasoning capabilities can be evaluated using GAIA \cite{mialon2023gaia}, while collaboration and communication skills can be assessed with RoCoBench \cite{mandi2024roco} particularly for embodied agents. 
\end{itemize}
However, publishing detailed agent profiles may inadvertently expose sensitive information about internal models, tools, or strategies. Therefore, developing privacy-preserving capability registration frameworks is critical to balance transparency with confidentiality.

\begin{figure}
\centering\setlength{\abovecaptionskip}{-0.08cm} \includegraphics[width=\linewidth]{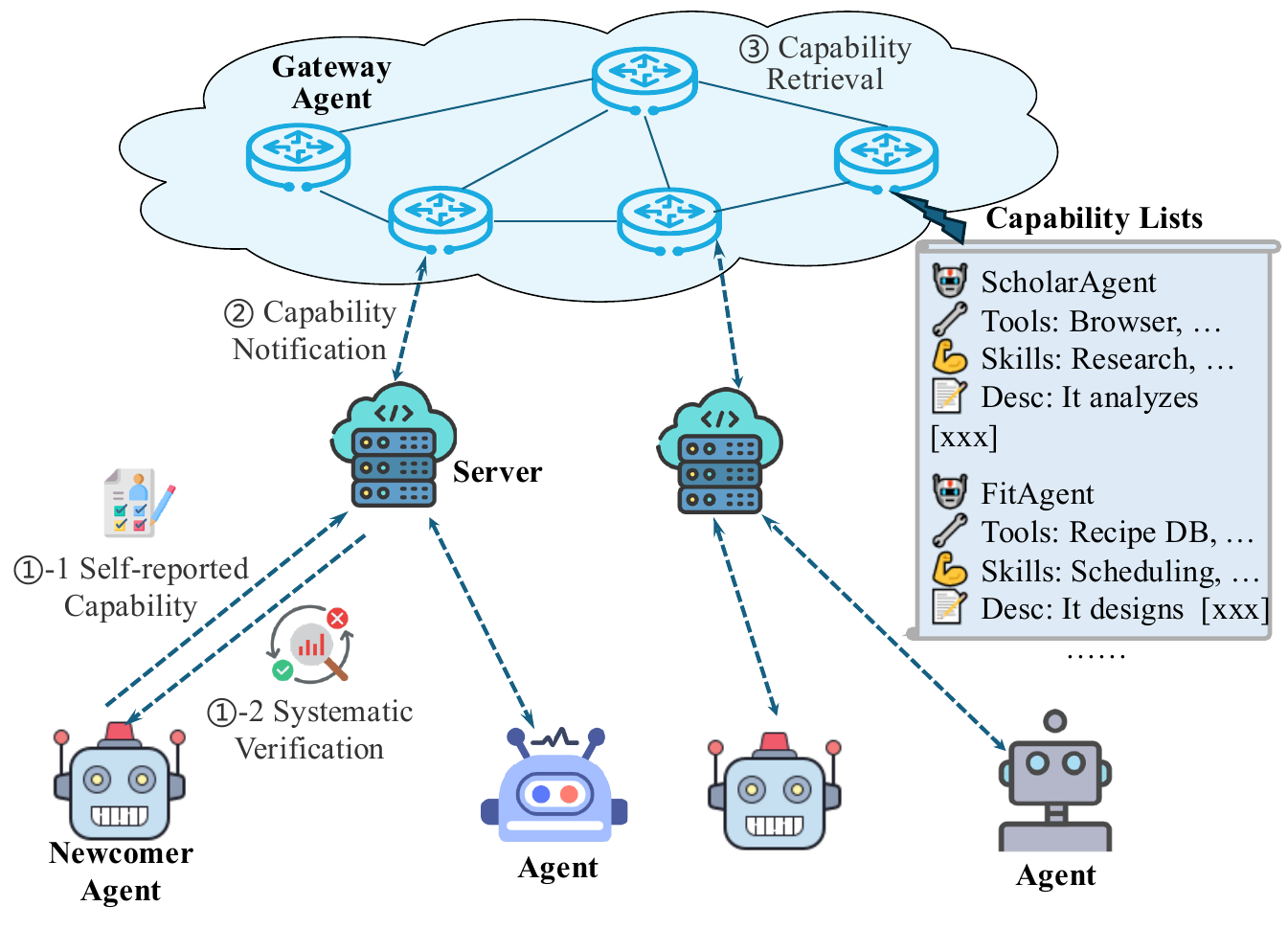}
    \caption{Illustration of capability notification and discovery in IoA.}
    \label{fig:Capability}
\end{figure}

 \subsubsection{Capability Notification of Agents}
Beyond static assessments, maintaining up-to-date knowledge of agent capabilities is crucial for real-time capability discovery and effective task orchestration in IoA systems. Given the dynamic evolution of large model-based agents, IoA should adopt capability notification strategies to ensure continuous synchronization between agents. 
\begin{itemize}
    \item \textit{Proactive Notifications.} This strategy involves agents autonomously notifying their gateway agents of any changes in their capabilities \cite{chen2025ioa}, such as the acquisition of new tools, skills, or model upgrades, to their gateway agents.
    \item \textit{Event-triggered Notifications.} In this mode, agent profiles are updated upon significant events, such as the successful completion of high-priority or complex tasks \cite{hu2021event}. By triggering updates only when necessary, this mechanism helps reduce unnecessary communication overhead compared to constant synchronization.
    \item \textit{Periodic Synchronization.} Complementing proactive and event-triggered strategies, periodic consistency checks at fixed intervals can be employed to maintain a reliable global view of agent capabilities across the system. 
\end{itemize}
However, the rapid evolution of agent capabilities introduces substantial communication overhead, posing significant challenges to the design of real-time synchronization mechanisms that balance freshness, scalability, and efficiency.

 \subsubsection{Capability Retrieval of Agents}
Efficient capability retrieval is crucial for identifying and selecting suitable agents whose capabilities align with specific task requirements, especially in large-scale and dynamic IoA. To accommodate varying levels of task complexity and agent heterogeneity, several retrieval strategies have been developed, including traditional search \cite{kraft2022fuzzy}, semantic retrieval \cite{bhopale2024transformer}, knowledge-based retrieval \cite{heuillet2022collective}, and agentic-enhanced retrieval \cite{singh2025agentic}.
\begin{itemize}
    \item \textit{Traditional Search.} These methods typically rely on exact or fuzzy matching \cite{kraft2022fuzzy} of agent profiles such as skill tags, toolsets, and domain expertise. While effective in well-structured scenarios with clearly labeled capabilities, they struggle to capture deeper semantic relationships, limiting their performance in complex or ambiguous queries.
    \item \textit{Semantic Retrieval.}  To capture richer semantics, semantic retrieval schemes leverage deep neural networks (DNNs), particularly transformer-based model (e.g., BERT), to encode agent profiles and queries into a shared embedding space \cite{bhopale2024transformer, zhang2025toward}. Such shared space captures the underlying semantic relationships between terms, allowing the retrieval system to understand the intent behind queries.
    \item \textit{Knowledge-based Retrieval.} This paradigm further improves retrieval accuracy and interpretability by integrating domain-specific knowledge into the retrieval process \cite{heuillet2022collective, zhang2025toward}. For instance, Heuillet \emph{et al.} \cite{heuillet2022collective} improve retrieval precision and reasoning robustness by fusing structured knowledge graphs with dense vector retrieval within a RAG framework. However, traditional RAG systems are limited by static workflows and lack the adaptability for multi-step reasoning and complex task management. 
\item \textit{Agentic-enhanced Retrieval.} To address the limitations of static RAG pipelines, recent approaches embed autonomous agents into the retrieval process \cite{singh2025agentic}. These RAG agents (or called agentic RAG) incorporate core capabilities such as reflection, planning, and tool use, enabling them to iteratively refine queries, dynamically select retrieval strategies, and interact with external tools or APIs as needed. This feedback-driven and adaptive retrieval process allows the system to better accommodate real-time task requirements, support context-aware reasoning. Consequently, agentic RAG frameworks offer significantly improved flexibility, scalability, and responsiveness for capability discovery in large and heterogeneous agent pools.
\end{itemize}

\begin{table*}[!t]
\centering\setlength{\abovecaptionskip}{0cm}
\caption{Comparison of Agent Capability Retrieval Strategies}\label{Table:Capability_Retrieval}
\begin{tabular}{ccccc}
\toprule
\textbf{Ref.}                               & \textbf{Retrieval Strategy} & \textbf{Key Technology}  & \textbf{Strengths}                                                                                                    & \textbf{Weaknesses}                                                                                            \\ \hline
\cite{kraft2022fuzzy}         & Traditional search          & Exact or fuzzy matching  & Simple, fast, computationally efficient                                                                               & Struggle with semantic ambiguity                                                                              \\ \hline
\cite{bhopale2024transformer} & Semantic retrieval          & DNNs (e.g., Transformer) & Capture intent and deeper semantics                                                                                  & \begin{tabular}[c]{@{}c@{}}Computationally heavy, \\ Require large annotated datasets\end{tabular}            \\ \hline
\cite{heuillet2022collective} & Knowledge-based retrieval   & Knowledge graphs, RAG  & Support interpretable and logical retrieval                                                                          & \begin{tabular}[c]{@{}c@{}}Expensive to build/maintain \\ Constrained by static workflow\end{tabular}         \\ \hline
\cite{singh2025agentic}       & Agentic-enhanced retrieval  & Agentic RAG              & \begin{tabular}[c]{@{}c@{}}Feedback-driven and adaptive retrieval, \\ Incorporate agents' capabilities\end{tabular} & \begin{tabular}[c]{@{}c@{}}Coordination complexity in IoA, \\ Require advanced AI infrastructure\end{tabular} \\ \bottomrule
\end{tabular}
\end{table*}

{\textit{Summary and Lessons Learned:}} 
Capability discovery between agents is a prerequisite for effective task allocation and collaborative orchestration in IoA, including three key components: capability evaluation, capability notification, and capability retrieval.
\ding{172} Capability evaluation includes self-reported declarations and system-level verification to construct agent capability profiles. While self-reporting facilitates fast registration, it may lead to inaccurate or exaggerated claims; system verification improves reliability but introduces scalability bottlenecks and potential privacy risks.
\ding{173} Capability notification maintains real-time system awareness through proactive updates, event-triggered mechanisms, and periodic synchronization. Nevertheless, frequent updates in large-scale IoA ecosystem can cause communication overhead and consistency issues, posing challenges to efficiency and scalability.
\ding{174} Capability retrieval via traditional matching, semantic retrieval, knowledge-enhanced methods, and agentic-enhanced strategies can improve the accuracy and flexibility of agent selection. Table~\ref{Table:Capability_Retrieval} provides a comparative summary of representative techniques and their key characteristics.

\subsection{{Interaction Structure \& Task Orchestration in IoA}}\label{subsec:Paradigm2}
To achieve optimal performance in the IoA, one of the core challenges lies in orchestrating agents to ensure efficient collaboration, which primarily involves two fundamental parts: \textit{interaction structure design} and \textit{task orchestration}.

\subsubsection{Interaction Structure Design}
The interaction structure design focuses on determining the optimal structure through which agents interact to execute tasks cooperatively. It encompasses two key elements: \textit{interaction mode} and \textit{communication topology}. The interaction mode defines how agents transmit, share, and update information with each other, which includes modes such as aggregate, reflect, or debate. Meanwhile, the communication topology specifies how agents are physically or logically connected within the network, with common structures including chain, tree, star, and graph-based configurations. 

\begin{figure}[!t]
\centering \setlength{\abovecaptionskip}{-0.cm}
\includegraphics[width=\linewidth]{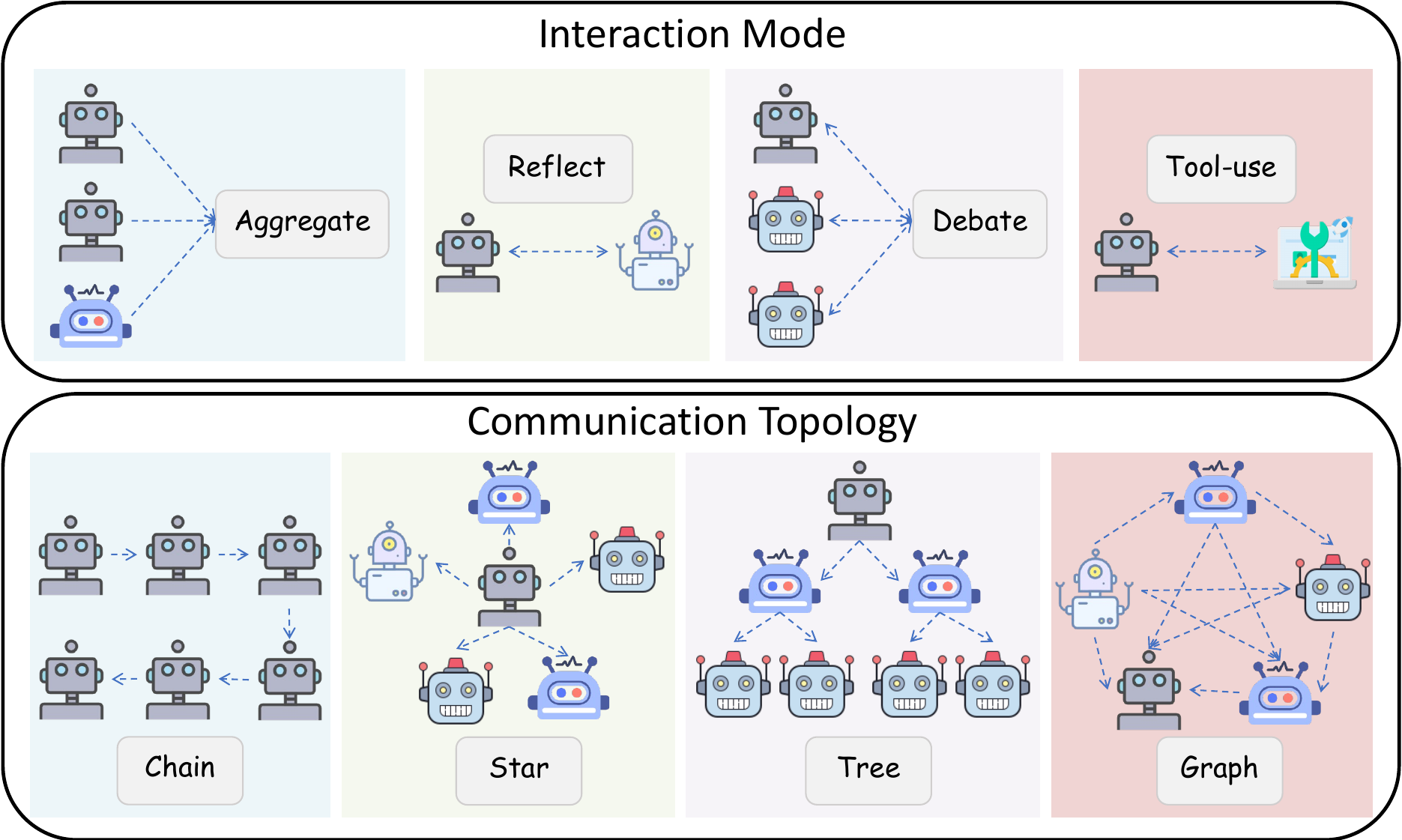}
\caption{Illustration of the commonly used interaction modes and communication topologies in IoA.}\label{fig:interaction&communication}\vspace{-2mm}
\end{figure}

\textit{a) Interaction Structure Selection.} Fig.~\ref{fig:interaction&communication} illustrates the commonly adopted interaction modes and corresponding communication topologies. 

\textit{\ding{172} Agent Interaction Modes.} Typical inter-agent interaction modes include aggregate, reflect, debate, and tool-use, as depicted in the upper part of Fig.~\ref{fig:interaction&communication}.

\begin{itemize}
    \item \textit{Aggregate.} In this mode, agents operate independently to generate diverse predictions, which are subsequently unified through an aggregation mechanism to yield a consensus output. Techniques such as majority voting~\cite{li2024more} and self-consistency analysis~\cite{chen2024universal} are widely utilized to enhance robustness and reliability in this collaborative mode.
    \item \textit{Reflect.} Agents function as reviewers, delivering feedback or suggestions based on previous outputs. This critical evaluation is iteratively incorporated into either the original generator or the feedback loop itself, leading to progressive refinement. Notable implementations include Self-Refine~\cite{madaan2023self} and Reflexion~\cite{shinn2023Reflexion}, which embody this reflective mode for agent performance enhancement.
    \item \textit{Debate.} This mode promotes argumentative collaboration, where agents engage in structured dialogue to challenge, justify, and improve upon each other's predictions. Each participant updates its response by considering the viewpoints and counterarguments presented by its peers. Empirical studies have shown that such multi-agent debates often yield more truthful and accurate conclusions compared to single-agent reasoning~\cite{du2023improving,liang2024encouraging}.
    \item \textit{Tool-use.} Tool-augmented agents can access external resources to improve task performance. This mode allows agents to invoke external functionalities such as RAG~\cite{lewis2020retrieval,li2024review} or executors for code validation via test cases~\cite{zhong2024debug,jiang2024ledex}, enabling a more informed and actionable response generation process. 
\end{itemize}

\textit{\ding{173} Topological Structures of Agent Networks.} In addition to interaction modes, communication topology plays a critical role in shaping the collaboration efficiency and scalability of multi-agent systems in IoA. As depicted in the lower part of Fig.~\ref{fig:interaction&communication}, typical topological structures include chain, star, tree, and graph, which are summarized as below.
\begin{itemize}
    \item \textit{Chain.} Agents are organized in a linear sequence, where each agent processes the input from its immediate predecessor and forwards the result to the next in line. This topology supports pipelined collaboration and has been employed in frameworks such as ChatDev~\cite{qian-etal-2024-chatdev}, MetaGPT~\cite{hong2024metagpt}, and L2MAC~\cite{holt2023l2mac}.
    \item \textit{Star.} A central agent, which is often functioning as a commander, coordinator, or teacher, oversees and guides a set of agents. This topology enables centralized control and is commonly used in systems such as AutoGen\cite{wu2024autogen}, SecurityBot\cite{yan2024depending}, and MiniGrid~\cite{zhou2024large}.
    \item \textit{Tree.} A hierarchical structure where a root agent supervises multiple child agents, which may in turn manage sub-agents, forming a multi-level control flow. The tree structure has been applied in frameworks such as SoA~\cite{ishibashi2024self}, enabling task delegation and modular problem solving.
    \item \textit{Graph.} A general communication structure where agents are interconnected in arbitrary patterns, including complete graphs, random graphs, or task-specific custom topologies~\cite{qian2024scaling,zhuge2024gptswarm}. This flexible structure supports P2P interaction and dynamic coordination among agents. 
\end{itemize}

\textit{b) Interaction Structure Optimization.}
To harness the potential of IoA systems, it is crucial to optimize both the interaction mode and the communication topology according to task characteristics. 

\textit{\ding{172} Interaction Mode Optimization.} Apart from directly adopting pre-defined interaction modes (e.g., aggregate and reflect) for multi-agent collaboration~\cite{zhang2024g,qian2024scaling,zhou2025multi}, recent research has extensively explored the optimization of interaction modes for enhanced collaboration among agents.
\begin{itemize}
    \item \textit{Debate-based Interaction.} One prominent line of work explores debate-based interactions. The MAD framework~\cite{du2023improving} introduces an asymmetric role design, where agents take on distinct roles such as debater and judge to evaluate arguments. In contrast, Liang \emph{et al.}~\cite{liang2024encouraging} propose a symmetric debate structure, enabling agents to engage in equal-status dialogues for collective decision-making. These paradigms have been successfully applied to tasks including machine translation and mathematical reasoning, yielding improved outcomes. Further refinements, such as ReConcile~\cite{chen2023reconcile}, leverage weighted voting across heterogeneous LLMs to enhance decision quality. 
    \item \textit{Role-based \& Cooperative Interaction.} Another direction focuses on role-based and cooperative interaction modes. The role-playing agent framework~\cite{li2023camel} enables agents to self-organize around predefined character roles, facilitating autonomous cooperation in complex scenarios. Similarly, Park \emph{et al.}~\cite{park2023generative} develop a sandbox environment with 25 virtual entities, each endowed with memory and personality traits, to simulate rich social behavior. Qian \emph{et al.}~\cite{qian-etal-2024-chatdev} utilize a chat-based software development framework, where agents assume software engineering roles to collaboratively complete full-stack development tasks at reduced costs. In addition to collaborative discussions, several works explore socially-aligned and self-collaborative paradigms. For instance, Liu \emph{et al.}~\cite{liu2023training} construct sandbox environments to curate behavior-aligned datasets and train LLMs with enhanced social alignment. Wang \emph{et al.}~\cite{wang2024unleashing} propose a self-collaboration method, where a single LLM is prompted with multi-persona roles, simulating multi-agent communication internally. This reduces communication latency while preserving the benefits of role diversity. 
    \item \textit{Multi-robot \& Embodied Agents.} Researchers have extended interaction optimization into multi-robot and embodied environments. Mandi \emph{et al.}~\cite{mandi2024roco} introduce a framework where multiple LLMs control individual robots, facilitating coordinated planning and execution. 
\end{itemize}

\textit{\ding{173} Communication Topology Optimization.} Optimizing the communication topology among agents is pivotal for improving the coordination and performance of IoA systems. Research in this area has evolved from early static designs to dynamic, task-adaptive, and learning-based approaches. Early explorations employed fully connected topologies, where each agent communicates with all others at every timestep \cite{foerster2016learning, sukhbaatar2016learning, peng2017multiagent}. Although straightforward, such dense connectivity becomes inefficient in complex environments due to excessive communication overhead and limited scalability. 
\begin{itemize}
    \item \textit{Prune Fully
Connected Topology.} Several methods have been proposed to prune unnecessary communications. For instance, IC3NET~\cite{singhlearning} and Gated-ACML~\cite{mao2020learning} introduce gating mechanisms that learn whether individual agents should participate in communication. SchedNet~\cite{kimlearning} learns a scheduler that selects a top-K subset of agents for each agent to interact with. ETCNET~\cite{hu2021event} penalizes communication cost during training to promote sparse connections. DC2Net~\cite{meng2023learning} decouples individual and group-level communication via two separate channels, further reducing overhead. 

\item \textit{Localized \& Observation-aware Topology.} Another line of work explores localized and observation-aware topologies. In ATOC~\cite{jiang2018learning}, a dynamically chosen initiator selects collaborators from its observation field. DGN~\cite{jianggraph} implements graph convolutional networks over locally observable agents, enabling structured information exchange. LSC~\cite{sheng2022learning} clusters agents within a predefined radius to form communication groups, promoting locality and scalability. 

\item \textit{Adaptive \& Task-driven Communication Structure.} More recent approaches focus on adaptive and task-driven communication structures. For instance, LLM-Blender\cite{jiang2023llm} orchestrates parallel LLM interactions using pairwise ranking, while task-specific partitioning~\cite{ning2023skeleton,qiao2024autoact} distributes workload and merges outputs via structural aggregation. Topological innovations such as linear stacking\cite{hao2025chatllm, zhang2023wider} and dynamic DAGs~\cite{zhang2023cumulative} have been applied for improved reasoning. To automate topology design, DyLAN\cite{liu2024dynamic} adaptively configures agent composition, Archon\cite{saad2024archon} frames collaboration as a hyperparameter optimization problem, and GPTSwarm\cite{zhuge2024gptswarm} employs RL to learn inter-agent connectivity. State-of-the-art systems such as ADAS\cite{hu2024automated} and AFlow~\cite{zhang2024aflow} use LLMs as controllers to explore topological variants through advanced search algorithms.
\end{itemize}

\begin{table*}[!t]
\centering\setlength{\abovecaptionskip}{0cm}
\caption{Comparison of existing task orchestration methods.}
\label{tab:task_assignment}
\begin{tabular}{c| >{\raggedright\arraybackslash}m{1.8cm} >{\raggedright\arraybackslash}m{4cm} >{\raggedright\arraybackslash}m{2.5cm} >{\raggedright\arraybackslash}m{3cm} >{\raggedright\arraybackslash}m{1.7cm}}
\toprule
 & \textbf{Strategy} & \textbf{Core Idea} & \textbf{Strengths} & \textbf{Weaknesses} & \textbf{Examples} \\
\midrule
\multirow{6}{*}{\makecell[c]{\textbf{Task} \\\textbf{Decomposition}}} & Rule-based Decomposition & Decomposing tasks based on pre-defined logical rules, schemes, or symbolic structures & High interpretability; Controllability & Limited adaptability to dynamic environments; High manual effort & TDAG~\cite{wang2025tdag}, HM-RAG~\cite{liu2025hm}\\
\cline{2-6}
& Learning-based Decomposition & Learning optimal task decomposition strategies through interaction with the environment using RL or LLMs & Strong adaptability; Automation without manual rule design & Higher training cost; Lower interpretability & HiggingGPT~\cite{shen2023hugginggpt}\\
\midrule
\multirow{6}{*}{\makecell[c]{\textbf{Task} \\\textbf{Allocation}}} & Routing-based Allocation & Dynamically matching task characteristics with suitable agents or models via optimized routing mechanisms & High efficiency, High scalability & Depends on the quality of routing strategy; Require additional optimization & RouteLLM~\cite{ong2406routellm}, Hybrid-LLM~\cite{dinghybrid}, RouterEval~\cite{huang2025routereval}\\
\cline{2-6}
& Self-organizing Allocation & Agents autonomously coordinate task and resource allocation through decentralized mechanisms such as artificial economies or RL & High autonomy; Reduced need for centralized control & Coordination overhead; Potential stability issues & Mindstorms~\cite{zhuge2023mindstorms}, Agora~\cite{marro2024scalable}\\
\bottomrule
\end{tabular}
\end{table*}

\subsubsection{Task Orchestration}
In IoA systems, effective task orchestration is essential for agents to operate collaboratively and efficiently. This process involves not only decomposing high-level requests into sub-tasks but also intelligently allocating these tasks based on agent capabilities, environmental context, and resource availability. Recent progress in task orchestration focuses on two key aspects: task decomposition and task allocation. 

\textit{a) Task Decomposition.} Task decomposition serves as the foundational step in task assignment within IoA systems, where high-level requests are broken down into manageable and executable sub-tasks. Recent advancements can be categorized into rule-based (or symbolic-based) decomposition and learning-based decomposition approaches. 

\begin{itemize}
    \item \textit{Rule-based decomposition.} Grounded in the classic divide-and-conquer paradigm~\cite{chendivide}, rule-based methods explicitly parse tasks according to pre-defined schemas or logical structures, enabling fine-grained control over complex objectives. For instance, TDAG~\cite{wang2025tdag} integrates task parsing with virtual sub-agent instantiation, achieving adaptive decomposition and flexible execution in multi-step real-world tasks. HM-RAG~\cite{liu2025hm} similarly decomposes complex queries into contextually coherent sub-tasks through semantic-aware query rewriting and schema-guided augmentation. While these approaches offer strong interpretability and control, they face challenges in dynamically changing environments.
 
    \item \textit{Learning-based decomposition.} In contrast, learning-based decomposition methods leverage environmental interactions to automatically infer optimal decomposition strategies. Shah \emph{et al.} propose a learning symbolic task decompositions framework~\cite{shah2025learning}, which simultaneously learns decomposition policies and sub-agent policies, enhancing sample efficiency and eliminating the need for manual rule design. In addition, the rise of LLMs has shifted decomposition into a data-driven paradigm. HuggingGPT~\cite{shen2023hugginggpt} exemplifies this shift by building a heterogeneous collaboration network that decomposes natural language tasks into multimodal subtasks, achieving a 19\% improvement in diagnostic accuracy.
\end{itemize}

\textit{b) Task Allocation.} After task decomposition, efficient task-agent assignments is crucial for effective task orchestration. Existing approaches can be broadly divided into routing-based allocation and self-organizing allocation strategies. 
\begin{itemize}
    \item \textit{Routing-based Allocation.} In routing-based allocation, systems dynamically select the most suitable agent or model for a given task by optimizing routing mechanisms. Frameworks such as RouteLLM~\cite{ong2406routellm} and Hybrid-LLM~\cite{dinghybrid} demonstrate the effectiveness of model selection routing, while RouterEval~\cite{huang2025routereval} provides a systematic evaluation protocol for routing quality. These methods offer high efficiency and scalability but are sensitive to the optimization quality of routing strategies.
 
    \item \textit{Self-organizing Allocation.} In contrast, self-organizing allocation strategies enable agents to autonomously coordinate resource usage and task distribution. Mindstorms~\cite{zhuge2023mindstorms} introduces an economy of minds (EOM) paradigm, where agents self-regulate computational resources based on the artificial economy framework with optimized cost-efficiency against task value. Agora~\cite{marro2024scalable} further explores RL to optimize communication and coordination among agents. Such decentralized methods are highly adaptive and eliminate the need for centralized control but introduce coordination overhead and convergence instability. Self-organizing allocation has been applied in various scenarios, including collaborative code generation~\cite{ishibashi2024self} and co-evolving reward-sharing mechanisms~\cite{ma2024coevolving}, highlighting its potential for scalable multi-agent collaboration.
\end{itemize}

{\textit{Summary and Lessons Learned:}} 
In IoA, well-designed interaction structures and task orchestration are foundational to achieve scalable and efficient multi-agent collaboration. On one hand, interaction structures combine both interaction modes and communication topologies. Interaction modes, such as aggregate, reflect, debate, and tool-use, define how agents share, refine, or challenge each other's outputs, while communication topologies, including chain, star, tree, and graph, shape efficiency, scalability, and coordination dynamics. Recent advances have moved beyond fixed designs, employing learning-based and task-driven approaches that adapt communication patterns and agent roles to contextual needs. On the other hand, task orchestration includes decomposing complex tasks and assigning subtasks to suitable agents. Rule-based and learning-based decomposition techniques offer complementary trade-offs between interpretability and adaptability, while routing-based and self-organizing task allocation strategies balance centralized efficiency and decentralized flexibility. Despite their diversity, these methods share a common goal: to enable dynamic, autonomous, and robust cooperation among heterogeneous agents in complex environments. Table~\ref{tab:task_assignment} summarizes representative orchestration approaches, highlighting their strategies, advantages, and limitations.

\begin{table*}[ht]
\centering\setlength{\abovecaptionskip}{0cm}
\caption{Comparison of Representative Agent Communication Protocols}
\label{tab:agent-protocols}
\begin{tabular}{cccccc}
\toprule
\textbf{Protocols}          & \textbf{MCP \cite{MCP}}& \textbf{A2A \cite{A2A}}& \textbf{ANP \cite{ANP}}& \textbf{AGNTCY \cite{AGNTCY}}& \textbf{Agora \cite{Agora}}\\ \hline
\textbf{Primary Goal}     & \begin{tabular}[c]{@{}c@{}}Large model's access\\to external resources\end{tabular}            & \begin{tabular}[c]{@{}c@{}}Agent-to-agent \\ communications\end{tabular}                           & \begin{tabular}[c]{@{}c@{}}Decentralized \\ networking\end{tabular}       & \begin{tabular}[c]{@{}c@{}}Standardized inter-agent \\ collaboration\end{tabular}        & \begin{tabular}[c]{@{}c@{}}Scalable LLM-agent \\ communications\end{tabular}                      \\ \hline
\textbf{Architecture}     & Client-Server                                                                      & P2P                                                                              & P2P                                                                     & Hybrid (Client-Server\,+\,P2P)                                                                       & P2P                                                                            \\ \hline
\textbf{Authentication}   & OAuth                                                                              & \begin{tabular}[c]{@{}c@{}}HTTP Auth, OAuth 2.0, \\API Keys, OpenID Connect\end{tabular} & \begin{tabular}[c]{@{}c@{}}W3C DID-based \\ decentralized\end{tabular}           & Agent connect protocol                                                                   & --                                                                                       \\ \hline
\textbf{Agent Discovery}  & --                                                                                 & Agent Cards                                                                               & Agent Description                                                                & OASF Metadata                                                                            & \begin{tabular}[c]{@{}c@{}}Natural language, \\ Protocol documents\end{tabular}          \\ \hline
\textbf{Endorsing Entity} & Anthropic                                                                          & Google                                                                                    & \begin{tabular}[c]{@{}c@{}}Open source\\  community\end{tabular}                 & \begin{tabular}[c]{@{}c@{}}Open source alliance \\ (e.g., Cisco, LangChain)\end{tabular} & \begin{tabular}[c]{@{}c@{}}Oxford University, \\ Camel AI\end{tabular}                   \\ \hline
\textbf{Limitations}      & \begin{tabular}[c]{@{}c@{}}Security risk, \\ Lack inter-agent focus \end{tabular} & \begin{tabular}[c]{@{}c@{}}No model \\ performance boost\end{tabular}                     & \begin{tabular}[c]{@{}c@{}}Inadequate privacy \\ and access control\end{tabular} & Conceptual design                                                                        & \begin{tabular}[c]{@{}c@{}}Limited validation \end{tabular}            \\ \hline
\textbf{Advantages}       & \begin{tabular}[c]{@{}c@{}}Open ecosystem, \\ Universal interfaces\end{tabular}    & \begin{tabular}[c]{@{}c@{}}Direct agent \\ collaboration\end{tabular}                     & Decentralized identity                                                           & Interoperability                                                                         & \begin{tabular}[c]{@{}c@{}}Balanced efficiency, \\ versatility, portability\end{tabular} \\ \bottomrule
\end{tabular}
\end{table*}

\begin{figure*}[htbp]
    \centering\setlength{\abovecaptionskip}{-0.05cm}
    \includegraphics[width= 0.66\textwidth]{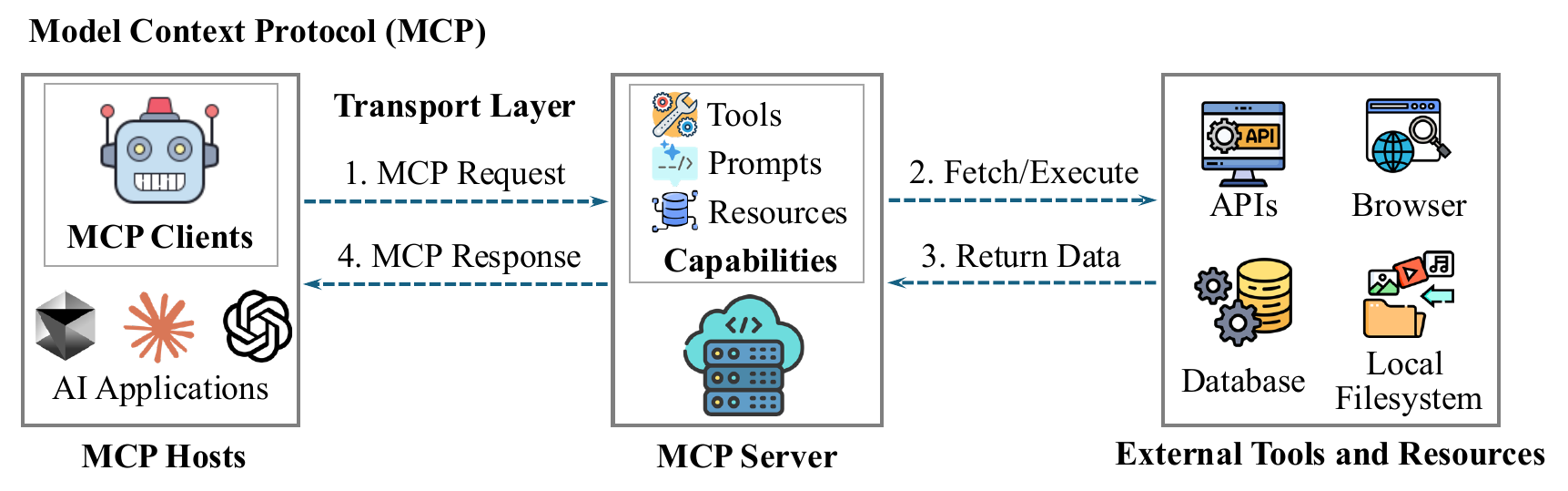}
    \caption{The workflow of Anthropic's MCP.}
    \label{fig:MCP}
\end{figure*}

\begin{figure*}[htbp]
    \centering\setlength{\abovecaptionskip}{-0.05cm}
    \includegraphics[width= 0.7\textwidth]{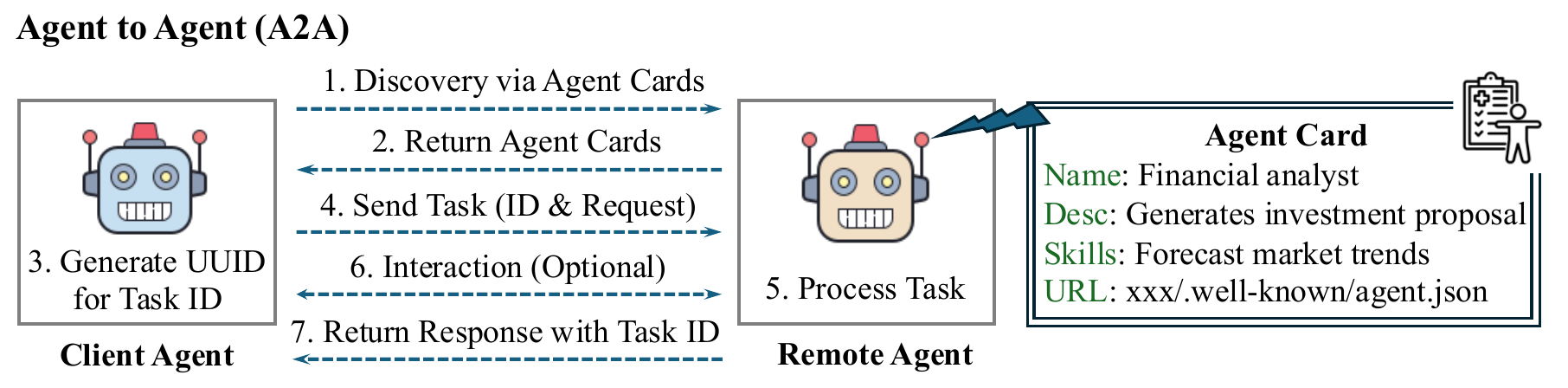}
    \caption{The workflow of Google's A2A.}
    \label{fig:A2A}
\end{figure*}

\subsection{Communication Protocols for IoA}\label{subsec:Paradigm3}
Traditionally, most agent frameworks only consider agents operating within their own ecosystems \cite{chen2025ioa,AutoGPT,wu2024autogen,LangChain,hong2024metagpt,BabyAGI,li2023camel}. Such isolated architecture hinders interoperability, restricts the integration of third-party agents, and fails to reflect real-world scenarios where agents are distributed across heterogeneous devices and contexts. 
To overcome these fragmentation challenges, recent efforts focus on standardized agent communication protocols to support message exchange, task negotiation, context sharing, identity management, and integration with external tools, forming the backbone of scalable and interoperable IoA ecosystems.
Representative agent communication protocols include the model context protocol (MCP) \cite{MCP}, agent-to-agent (A2A) \cite{A2A}, agent network protocol (ANP) \cite{ANP}, AGNTCY \cite{AGNTCY}, and Agora \cite{Agora}, as summarized in Table~\ref{tab:agent-protocols}.

\subsubsection{Anthropic's MCP}
The MCP, initiated by Anthropic, is an open standard designed to enable seamless integration between LLMs and external tools, data sources, and systems. By providing a unified interface and leveraging authorization via OAuth \cite{hou2025model}, MCP allows agents to access real-time data and services without exposing user credentials, thereby enhancing contextual awareness and response relevance.

As shown in Fig.~\ref{fig:MCP}, MCP adopts a client-server architecture. The host (e.g., Claude desktop) runs an MCP client that connects to one or multiple MCP servers \cite{MCP}. Each server exposes specific capabilities, categorized as tools (i.e., executable functions), resources (e.g., data sources), or predefined prompts. A typical interaction begins with protocol negotiation and capability discovery between the MCP client and server. The client sends an initial request to retrieve the server’s available capabilities, and the server responds accordingly. When the client requires a specific function (e.g., calling external APIs), it sends a request to the appropriate MCP server. The server executes the corresponding tool or retrieves the relevant resource, then returns the result. The host integrates this output into the ongoing interaction, enabling more accurate and context-aware responses. By standardizing this communication pipeline, MCP decouples model behavior from tool implementation, supporting modular, extensible, and cross-platform agentic workflows.

\subsubsection{Google's A2A}
Unlike the MCP, which focuses on connecting individual LLMs with external data sources and tools, Google's A2A protocol serves as a networking layer to facilitate seamless interoperability among heterogeneous AI agents. It facilitates direct communication and knowledge sharing for task collaboration.
A2A incorporates enterprise-grade authentication and authorization mechanisms compliant with OpenAPI standards by default \cite{A2A}, such as OAuth 2.0 and OpenID Connect.

A2A facilitates communication between a client agent and a remote agent \cite{A2A}, as shown in Fig.~\ref{fig:A2A}. The client agent formulates and transmits tasks, while the remote agent responds by executing the tasks or providing relevant information. Agent collaboration begins when the client agent discovers a remote agent by retrieving its agent card, i.e., a standardized JSON file typically hosted at a well-known URL. This agent card contains essential metadata, including agent's capabilities, skills, endpoint address, and authentication requirements, enabling structured agents discovery across diverse systems. Then, the client agent initiates a task by sending a request containing a unique task ID via universally unique identifier (UUID).The remote agent handles the task according to its declared capabilities. For asynchronous tasks, intermediate server-sent events (SSE) are streamed as the task progresses, providing status updates and partial response. In contrast, synchronous tasks return the final result directly in the response. If additional input is needed, the client sends follow-up messages using the same task ID until the task reaches a terminal state (e.g., completed or failed).

\subsubsection{Other Protocols}
In addition to MCP and A2A, several emerging communication protocols are designed for IoA systems. 
\begin{itemize}
    \item \textit{ANP.} The ANP protocol \cite{ANP} defines a fully decentralized P2P architecture in which each agent holds a W3C DID \cite{halpin2020vision} for mutual authentication and uses end-to-end encryption to secure all messages. It also supports dynamic protocol negotiation to enable flexible and efficient collaboration.
    \item \textit{AGNTCY.} The AGNTCY protocol \cite{AGNTCY} focuses on building an open and interoperable IoA infrastructure. Built atop the agent connect protocol \cite{ACP}, it provides a standard interface for agent invocation and specifies agent capabilities and communication patterns, facilitating standardized inter-agent collaboration.
    \item \textit{Agora.} As a research project, Agora \cite{Agora} develops a scalable agent communication protocol for large-scale IoA deployments. It employs standardized routines for frequent communication and leverages natural language or LLM-generated routines for less frequent interactions, aiming to balance versatility, efficiency, and portability in agent communications \cite{Agora}.
    \item {\textit{Mixture-of-Agents (MoA).} MoA \cite{wang2024mixture} enhances agent coordination emulating a ``collaborative LLM'' through a layered architecture, where each layer contains multiple LLM-based agents with the same model architecture that refine the previous layer's outputs as auxiliary context. 
    }
\end{itemize}

{\textit{Summary and Lessons Learned:}} 
In IoA systems, standardized communication protocols are vital to enable interoperability among heterogeneous agents, laying the foundation for scalable and context-aware collaboration. Notably, MCP provides agents with structured and standardized access to external tools and resources through unified interfaces, while A2A emphasizes P2P interactions and dynamic agent discovery. In parallel, ANP focuses on decentralized networking; AGNTCY adopts a hybrid architecture to promote standardized inter-agent collaboration; Agora explores agent integration with a balance of efficiency, versatility, and portability; {and MoA targets on task-oriented layered agent orchestration.} Despite differences in architecture and functionality, these protocols share a common goal: to build thriving, extensible, and trustworthy IoA ecosystems. Table~\ref{tab:agent-protocols} provides a comparative summary of their core goals, architectural designs, authentication mechanisms, agent discovery strategies, endorsing entities, advantages, and limitations.

\subsection{Consensus \& Conflict Resolution in IoA}\label{subsec:Paradigm4}
Consensus in IoA refers to multiple agents achieving agreement on states, decision, or actions. Although traditional MAS offer plenty of consensus methods \cite{amirkhani2022consensus}, they often fall short in LLM-enabled IoA due to its unique characteristics, including sophisticated interactions, diverse reasoning patterns, large-scale structure. Specifically, i) agent communications in IoA is highly sophisticated \cite{chen2025ioa}, requiring mechanism for turn-taking, coherence maintenance, and conflict resolution. ii) LLM agents exhibit diverse reasoning behaviors \cite{li2024survey}, often producing inconsistent or hallucinated outputs, which complicates consensus formation. iii) IoA systems typically operate at a large scale, where the growing number of agents demands scalable and efficient consensus mechanisms. In response, the following subsections present representative approaches designed to ensure coordination, reliability, and scalability in IoA consensus.

\textit{1) How to Regulate Turn-Taking and Conflicts?}
To manage communication conflicts and maintain coherence, IoA systems require sequential speaking mechanisms that regulate speaking turns during group collaboration \cite{chen2025ioa,holt2023l2mac}.
\begin{itemize}
    \item \textit{Polling Mechanism}: A coordinator sequentially queries agents for their states or opinions, enforcing only one agent speaks at a time to reduce collisions. For instance, in \cite{chen2025ioa}, Chen \emph{et al.} model the conversation flow as a finite state machine with five predefined state, where LLMs autonomously manage state transitions to coordinate collaboration. Although effective for maintaining order, it introduces latency and may become a bottleneck in large-scale deployments.
    \item \textit{Arbitration Control}: When conflicts arise, a third-party arbitrator (e.g., higher-tie agent or rule engine) can resolve contentions and assigns speaking rights or priorities, ensuring conflict resolution across agents. For instance, the role arbitration framework proposed by Franceschi \emph{et al.} \cite{franceschi2023human} , which models leader-follower dynamics via differential game theory, holds potential for adaptation to IoA settings. By dynamically assigning speaking rights or decision-making priorities based on task context and agent capabilities, such mechanisms can facilitate smoother coordination and effective conflict resolution in complex multi-agent environments.

\end{itemize}

\textit{2) How to Align Diverse Agent Reasoning?}
To overcome the inherent limitations of LLM agents, such as diverse reasoning paths and hallucination, reasoning alignment techniques extend beyond simple majority voting \cite{li2024more} to achieve more reliable consensus. {Semantic interoperability remains a key challenge for the IoA, particularly as agents become increasingly heterogeneous in model architecture, training data, and tool usage. Such diversity often leads to issues such as sycophancy (agents reinforcing each other’s responses without critical reasoning), prompt ambiguity, and difficulty in reaching robust consensus.}
\begin{itemize}
    \item \textit{Self-Consistency Verification}: Individual agents could perform multiple independent reasoning traces on the same query and cross-validates the outputs. By checking the internal consistency across these traces, agents can filter out hallucinated or unstable responses \cite{chen2024universal}.
    
    \item {\textit{Dynamic Prompt Optimization}: To improve alignment among LLM-based agents, CONSENSAGENT \cite{pitreconsensagent} introduces a dynamic prompt optimization mechanism to refine prompts based on agent interactions, which mitigates sycophancy and improves consensus efficiency in multi-agent LLM debates.  }
    
    \item \textit{Collective Reasoning}: Beyond prompt engineering, another promising direction is to utilize the collective power of multiple LLM agents to enhance semantic alignment, enabling mutual verification and augmentation. Such collective process enhances inference depth and robustness, particularly for complex, multi-hop reasoning tasks. 
    For instance, He \emph{et al.} \cite{he2022wisdom} demonstrate that aggregating reasoning from diverse models can improve factual accuracy and approximate correctness, even without ground truth.
    
    {\textit{a) Round-table discussion.} In \cite{chen2023reconcile}, the RECONCILE scheme leverages round-table discussions among diverse LLM agents, incorporating confidence-weighted voting and convincingness to enhance collaborative reasoning and consensus-building.} 
    
    {\textit{b) Multi-agent deliberation.} To further improve trustworthy negotiation, Yang \textit{et al.} \cite{yang2024confidence} present a training-free, multi-agent deliberation protocol where agents collectively calibrate their confidence estimates through argumentation, peer feedback, and majority voting to calibrate agent confidence and improve reliability in collective decision-making.}
    
    {\textit{c) MoA.} To address the challenge of aligning agent outputs to specific task requirements at inference time, Chakraborty \textit{et al.} \cite{chakraborty2025collab}  propose an MoA controlled decoding strategy, enabling inference-time alignment by dynamically selecting the most suitable agent for each token, thus facilitating fine-grained semantic interoperability without retraining.} 
    
    \item {\textit{Ontology Matching:} For more structured semantic alignment tasks, such as ontology matching, Qiang \textit{et al.} \cite{qiang2025agentom} propose an agent-powered LLM framework named Agent-OM that uses Siamese LLM agents to retrieve, match, and validate entity correspondences across heterogeneous ontologies, employing CoT reasoning and RAG to achieve semantic alignment between heterogeneous knowledge sources.}
\end{itemize}

\textit{3) How to Scale Consensus in Large-Scale IoA?}
To mitigate scalability issues in consensus formation, hierarchical consensus mechanisms could be employed. Agents are partitioned into local clusters based on network topology or task similarity, with each cluster forms local consensus independently. Then, cluster representatives propagate aggregated results to higher-level coordinators, enabling efficient, scalable, and coherent system-wide decision-making.
For instance, Feng \emph{et al.} \cite{feng2024hierarchical} propose hierarchical consensus framework that forms global consensus from agents' local observations via contrastive learning. By incorporating multi-layer consensus and an adaptive attention mechanism, the framework enhances decentralized cooperation and improves scalability. 

{\textit{Summary and Lessons Learned:}} 
Consensus and conflict-resolution mechanisms are essential for coordinated decision-making in IoA, especially under complex interaction and workflow orchestration. It comprises two core elements: communication regulation and reasoning alignment.  
Communication regulation ensures ordered interaction through mechanisms such as polling-based turn-taking and arbitration control. 
However, it faces bottleneck-related challenges in centralized settings, or suffer from high latency in distributed settings.
Moreover, reasoning alignment mitigates inconsistency and hallucination via (i) self-consistency checks, which cross-validate multiple inference traces to enhance individual agent reliability, and (ii) collective reasoning, where diverse agent outputs are mutually verified to strengthen the robustness of multi-step inference. 
Additionally, in large-scale IoA systems, scalable consensus mechanisms employ hierarchical or clustered architectures to maintain efficiency. By achieving local consensus within clusters and aggregating results via higher-level coordinator agents, thereby reducing communication overhead while ensuring coherent system-wide decisions.

\subsection{Economic Models of IoA}\label{subsec:Paradigm5}
Well-established economic models can foster long-term healthy, fair, and active agent collaborations in large-scale decentralized IoA ecosystems. By implementing dynamic pricing strategies and incentive \& penalty mechanisms, agents are motivated to participate honestly and proactively \cite{geng2024noncooperative, 10704033}, thereby suppressing selfish behaviors and inactive participation.

\subsubsection{Pricing Models}
One key aspect of economic models in IoA systems is pricing mechanisms, which typically fall into two paradigms: capability-based pricing \cite{duetting2024mechanism, bergemann2025economics} and contribution-aware pricing \cite{heuillet2022collective}. 
\begin{itemize}
    \item \textit{Capability-based Pricing} charges agents according to their consumption of underlying large model resources, where common metrics include token usage, number of interactions, and model complexity, thereby aligning costs with the computational load each agent imposes \cite{duetting2024mechanism}.
    \item \textit{Contribution-aware Pricing} rewards agents based on their marginal impact on task outcomes. A common approach employs the Shapley value to quantify each agent's marginal contribution. Specifically, agents with greater influence over critical task components receive proportionally larger shares of the collective reward, enabling fair profit distribution in multi-agent settings. For instance, Heuillet \emph{et al.} \cite{heuillet2022collective} apply Monte Carlo sampling to estimate Shapley values in a MARL setting to infer and reward cooperative behavior in complex environments.
\end{itemize}

\subsubsection{Economic Mechanism Design}
Economic incentive and penalty mechanisms are essential for promoting sustained participation and cooperative behavior in IoA ecosystems.

\begin{table*}[ht]
\centering\setlength{\abovecaptionskip}{0cm}
\caption{Comparison of Incentive Mechanism Designs in IoA Systems}\label{tab:Incentive_mechanism}
\begin{tabular}{ccccc}
\toprule
\textbf{Ref.}                               & \textbf{Mechanism Type}                                               & \textbf{Advantages}                                                                                         & \textbf{Limitations}                                                                            & \textbf{Key Methodology}                                                                                        \\ \hline
\cite{you2024privacy}         & Auction theory                                                        & Fairness, efficiency, decentralization                                                                      & \begin{tabular}[c]{@{}c@{}}High computation, require \\ accurate valuation models\end{tabular} & \begin{tabular}[c]{@{}c@{}}Privacy-preserving stochastic auction\\with multi-agent competition\end{tabular}   \\ \hline
\cite{ye2024optimizing}       & Contract theory                                                       & \begin{tabular}[c]{@{}c@{}}Mitigate information asymmetry, \\promote long-term cooperation\end{tabular}  & \begin{tabular}[c]{@{}c@{}}Complex design, inflexible\\ in dynamic environments\end{tabular}   & \begin{tabular}[c]{@{}c@{}}Multi-dimensional contract optimization \\ under asymmetric information\end{tabular} \\ \hline
\cite{geng2024noncooperative} & \begin{tabular}[c]{@{}c@{}}Non-cooperative\\ game\end{tabular}       & \begin{tabular}[c]{@{}c@{}}Reflect realistic decision-making,\\support distributed control\end{tabular} & \begin{tabular}[c]{@{}c@{}}Potential sub-optimal \\ global outcomes\end{tabular}              & \begin{tabular}[c]{@{}c@{}}Generalized NE modeling under\\physical \& comm. constraints\end{tabular} \\ \hline
\cite{10704033}               & Cooperative game                                                      & \begin{tabular}[c]{@{}c@{}}Ensure fairness, improve \\ system performance\end{tabular}            & \begin{tabular}[c]{@{}c@{}}Coalition instability,\\ free-riding risk\end{tabular}               & \begin{tabular}[c]{@{}c@{}}Coalition formation with Shapley \\ value-based benefit distribution\end{tabular}    \\ \hline
\cite{xu2024semantic}         & \begin{tabular}[c]{@{}c@{}}Reputation \\ mechanism\end{tabular} & Support sustainable collaboration                                                                          & \begin{tabular}[c]{@{}c@{}}Prone to manipulation,\\ require monitoring\end{tabular}      & \begin{tabular}[c]{@{}c@{}}Hierarchical worker selection with \\ semantic-aware reputation scoring\end{tabular} \\ \bottomrule
\end{tabular}
\end{table*}

\textit{a) Incentive Mechanism Design.} Typically, agents are rewarded through monetary rewards, e.g., tradable tokens redeemable for computing resources, based on task performance \cite{bergemann2025economics}. 
{Freni \textit{et al.} \cite{FRENI2022100069} present a comprehensive model for agent-driven token economy design which integrates token-based incentives and governance tokens to align agent interests, support decentralized decision-making, and promote ecosystem sustainability.} 
Prior works on incentives in MAS provides valuable lessons for IoA, including auction theory \cite{you2024privacy}, contract theory \cite{ye2024optimizing}, game theory \cite{geng2024noncooperative, 10704033}, and reputation-based mechanisms \cite{xu2024semantic}.
    \begin{itemize}
        \item \textit{Auction Theory.} In auction mechanisms, agents bid for tasks based on resource availability and task valuation, enabling fair competition and efficient allocation. For instance, {Yang \textit{et al.} \cite{yang2025agent} introduce the concept of an agent-centric economy and propose Agent Exchange (AEX), i.e., a real-time auction platform where agents act as autonomous economic actors, bidding and negotiating for tasks and resources. The AEX framework leverages market-based protocols and dynamic value attribution (e.g., Shapley value) to ensure fair and efficient agent coordination.} You \emph{et al.} \cite{you2024privacy} further design a stochastic auction mechanism for efficient and privacy-preserving resource allocation among multiple agents. 
        \item \textit{Contract Theory.} By specifying performance-based contract items for agents with diverse types, contract-theoretic approaches provide a structured approach to mitigate information asymmetry and incentivize agents to align their actions with the objective of the principal (e.g., task coordinator agent). For instance, Ye \emph{et al.} \cite{ye2024optimizing} employ contract theory and generative diffusion models to optimize multi-dimensional resource allocation under asymmetric information, offering insights for incentive design in IoA systems. {Besides, Ivanov \textit{et al.} \cite{ivanov2024principalagent} combine RL with principal-agent theory, illustrating that contract-based incentives can align agent behavior with system-level objectives in sequential and multi-agent environments.}
        \item \textit{Game Theory.} Game-theoretic approaches in IoA are divided into non-cooperative and cooperative models games. \textit{\ding{172} Non-cooperative games} model agents as rational and selfish entities whose strategic interactions converge to a Nash equilibrium, at which no agent can unilaterally improve its payoff by deviating. For instance, Geng \emph{et al.} \cite{geng2024noncooperative} formulate a non-cooperative game among distributed quadrotor UAVs and derive a generalized Nash equilibrium of the game under communication and physical constraints.
        \textit{\ding{173} Cooperative games} encourage agents to independently form coalitions, with coalitional benefits distributed fairly among its members via mechanisms such as Shapley values. For instance, Wang \emph{et al.} \cite{10704033} model social cluster formation as a cooperative game, where agents collaboratively optimize group strategies to enhance overall utility while ensuring equitable coalition benefit allocation. {From a joint behavioral and game-theoretic perspective, Lalmohammed  \cite{mohammed2025welfare} develops a welfare modeling framework for human-AI economic ecosystems, incorporating trust evolution, risk aversion, and collaboration synergies. Using agent-based modeling, the study shows that dynamic trust-building and equitable incentive distribution are pivotal for maximizing collective welfare in mixed human-AI systems.}
        \item \textit{Reputation mechanisms} track agents' historical performance, prioritizing high-reputation agents for critical tasks and discouraging malicious behavior. For instance, Xu \emph{et al.} \cite{xu2024semantic} propose a hierarchical reputation framework that selects reliable agents based on reputation scores for secure collaboration in dynamic and uncertain environments. {To further enable transparent and automated incentive mechanisms, Karim \textit{et al.} \cite{karim2025ai} explore the integration of AI agents with blockchain, showing how smart contracts and decentralized ledgers facilitate agent-to-agent payments, reputation, and governance.}
    \end{itemize}
    
\textit{b) Penalty Mechanism Design.} 
Effective penalty mechanisms can deter free-riding and malicious behaviors by imposing costs (e.g., token deductions or reputation losses) on agents that fail to complete tasks such as missing deadlines or generating low-quality or hallucinated responses \cite{wang2024large}. Key penalty approaches include:
\begin{itemize}
    \item \textit{Game-theoretic Penalties.} Leveraging principles from game theory, several incentive-compatible penalty schemes have developed to align individual agent behavior with system-wide objectives. For instance, Xu \emph{et al.} \cite{8387487} design game-theoretical reward and penalty policies based on agents' reported beliefs and past performance, thereby improving task quality while reducing latency.
    \item \textit{Optimization-based Penalties.} From the optimization perspective, penalties can be incorporated into fitness evaluation functions to guide decentralized agents toward global goals while respecting local constraints. For instance, Chen \emph{et al.} \cite{10500484} introduce a conflict-aware penalty function in a multi-agent co-evolutionary optimization algorithm, effectively coordinating distributed agents with limited information under practical constraints. Simulation results show that this approach effectively mitigates conflicts and ensures solution feasibility.
    \item \textit{Blockchain-based Penalties.} Blockchain systems can be integrated into IoA systems to enable transparent, tamper-resistant enforcement of penalties. For instance, Motepalli \emph{et al.} \cite{9569791} propose slashing strategies within a blockchain framework to automatically penalize free-riders via token forfeiture or reputation reduction, thereby bolstering trust and accountability.
\end{itemize}

{\textit{Summary and Lessons Learned:}} 
Economic models in IoA align individual agent incentives with collective objectives to sustain long-term collaboration. 
In IoA pricing models, capability-based pricing charges agents according to resource usage (e.g., token usage), whereas contribution-aware pricing allocates rewards agents proportionally to their marginal contributions. However, accurately estimating contributions in large-scale and dynamic IoA environments is often computationally prohibitive.
Beyond pricing, mechanism design integrates incentives and penalties to foster cooperation. Incentive mechanisms grounded in auction, contract, and game theories promote active participation, while reputation mechanisms support long-term collaboration. Table~\ref{tab:Incentive_mechanism} summarizes representative incentive works on agents along with their strengths and limitations. Conversely, penalty mechanisms deter misbehaving agents by imposing costs (e.g., token forfeiture or reputation loss). A key challenge is developing interoperable and fair economic frameworks across heterogeneous IoA platforms with varying protocols and value models.

\subsection{Trustworthy Regulation in IoA}\label{subsec:Paradigm6}
As AI agents gain increasing autonomy and direct influence over both digital and physical environments, ensuring trust, safety, and ethical compliance in the IoA becomes critical. Unlike conventional IoT devices, agents can autonomously make decisions, initiate actions, and interact with other agents in complex ways, yet they lack legal accountability. Therefore, IoA requires trustworthy regulatory mechanisms that can verify agent identities, govern their behaviors, and safeguard against malicious activities. Potential IoA regulation technologies include DIDs, blockchain, and legal regulations, as below. 

{\textit{1) Accountability of AI Agents.}
Accountability in agentic systems requires robust mechanisms for tracing, attributing, and enforcing responsibility across the agent lifecycle. For instance, Chan \textit{et al.} \cite{chan2024visibility} propose a multi-layered visibility framework, including agent identifiers, real-time monitoring, and activity logging, to enable traceability and post-incident forensics. These measures facilitate the identification of responsible parties, including developers, deployers, or users, when agents cause harm or violate policies. Tamang and Bora \cite{tamang2025enforcement} introduce the Enforcement Agent (EA) framework, which embeds supervisory agents within multi-agent environments to monitor peer behavior, detect misalignment, and intervene in real time. The results demonstrate that such embedded oversight significantly enhances system safety and distributed accountability, especially in adversarial or dynamic settings. As reported by \cite{AheadoftheCurve}, the “many hands problem” in agent value chains emphasizes the need for clear allocation of accountability among model providers, system providers, and deployers.}
\begin{itemize}
    \item \textit{DID and Verifiable Credentials (VCs).} DID frameworks (e.g., W3C DID \cite{halpin2020vision} and Sovrin \cite{9582551}) assign agents self-sovereign identifiers, which are independent of centralized authorities and anchored in tamper-resistant ledgers \cite{mazzocca2025survey}. With VCs, agents can issue, present, and revoke cryptographically signed credentials such as trust ratings, compliance certificates, or capability attestations in a privacy-friendly manner (i.e., without disclosing underlying data) \cite{mazzocca2025survey}. Modern DID implementations should support on-demand DID creation, context-specific key rotation, and selective disclosure proofs, to enable privacy-preserving attribute validation across domains.
    However, in time-varying IoA environments, maintaining up-to-date DID documents and revocation statuses at large-scale IoA needs near-instant ledger updates and efficient distributed resolution services, which requires further investigations.
     
    \item \textit{Blockchain and Distributed Ledger Technologies.} Blockchain (e.g., Hyperledger Fabric and Ethereum) offers an immutable and decentralized infrastructure for managing digital identities, recording agent actions, and enabling transparent auditing \cite{9569791,calvaresi2018multi}. Smart contracts can enforce predefined policies for agent authorization, behavior verification, accountability, and decentralized governance in a trustless environment \cite{9632411}. Despite its strengths, blockchain suffers from latency, limited throughput, and high energy consumption issues that conflict with real-time coordination in large-scale IoA. Moreover, most blockchains are not designed for privacy-preserving operations \cite{bernabe2019privacy}, which may expose sensitive agent behavior or user data. Emerging solutions such as sharding, off-chain channels, and zero-knowledge proofs mitigate these issues but also add complexity to protocol design and interoperability \cite{9631953}.
\end{itemize}

{\textit{2) Regulatory Compliance.}} As agents assume decision-making roles in critical domains, comprehensive legal frameworks should define agent owner, liability regimes, and oversight procedures. Regulatory sandboxes \cite{zetzsche2017regulating} and certification processes can vet agent behaviors in controlled environments, while dynamic rule frameworks (combining statutes, industry standards, and ethical guidelines) set operational boundaries. Algorithmic audits and mandated explainability reports further ensure traceability and compliance.
{Besides, regulatory compliance for AI agents is increasingly defined by regulations such as the EU AI Act and GDPR. For instance, a comprehensive analysis of the EU AI Act's application to AI agents is provided in \cite{AheadoftheCurve}, which outlines four key governance pillars: risk assessment, transparency tools, technical deployment controls, and human oversight. It further specifies how compliance obligations, such as risk identification, agent identification, activity logging, and emergency shutdowns, should be allocated across the agent value chain. Complementing this, Chan \textit{et al.} \cite{chan2024visibility} highlight the importance of technical measures for data traceability, privacy protection, and auditability to support regulatory investigations and uphold user rights under GDPR.}

Global IoA deployments face jurisdictional fragmentation and conflicting regulations, necessitating harmonized cross-border enforcement protocols and data-sovereignty agreements. Moreover, apportioning liability in multi-agent collaborations, where agents autonomously negotiate and delegate tasks, requires novel legal constructs to assign fault among developers, operators, and the agents themselves, without stifling innovation.

{\textit{3) Ethical and Responsible AI Agents.} Embedding ethical reasoning and value alignment is essential for responsible AI autonomy. Pujari \textit{et al.} \cite{tejaskumarpujari2024ethical} advocate for integrating ethical principles, such as virtue ethics, deontology, and consequentialism, into agent decision-making, and highlight the need for norm-aware, transparent behavior and adaptive governance. Hu \textit{et al.} \cite{hu2025trustless} further examine the challenges of self-sovereign decentralized agents (DeAgents) on trustless infrastructures, emphasizing the importance of participatory governance, flexible protocols, and robust fail-safes to ensure alignment with societal values even in decentralized and permissionless environments.}

{\textit{Summary and Lessons Learned:}} 
Ensuring trusted regulation in IoA systems requires both technical safeguards and legal frameworks. On the technical side, DIDs, VCs, and blockchain enable secure agent identification, behavior verification, and transparent auditing. However, scalability, privacy, and interoperability challenges remain. Legally, a combination of hard law (e.g., laws and regulations) and soft law (e.g., community norms and ethical standards) is essential to regulate autonomous agent behavior, allocate responsibility, and support cross-border governance. 

\section{Future Research Directions}\label{sec:FUTUREWORK}
In this section, we point out several future directions to be investigated in the era of IoA from the following aspects.

{\subsection{IoA Standardization and Interoperability Frameworks}\label{subsec:future0}
The widespread adoption of IoA systems necessitates unified standardization frameworks to ensure cross-platform and cross-domain interoperability. Current standardization efforts are fragmented across multiple organizations. For instance, IEEE SA-P3394\footnote{https://standards.ieee.org/ieee/3394/11377/} focuses on LLM agent interface specifications, while ITU-T F.748.46\footnote{https://www.itu.int/rec/T-REC-F.748.46-202503-I/en} addresses AI agent requirements and evaluation methods. 
A critical challenge is the absence of unified protocols for agent discovery, capability negotiation, and dynamic service composition across heterogeneous platforms. Current approaches rely on proprietary interfaces that limit cross-platform collaboration and create vendor lock-in scenarios. Additionally, inconsistent security models and authentication mechanisms hinder trust establishment between agents from different organizations or domains. Future research should prioritize developing standardized agent description languages and lightweight communication protocols that enable seamless integration while preserving autonomy.}

\subsection{Secure \& Adaptive Agent Communication Protocols}\label{subsec:future1}
As IoA systems scale across domains, from virtual assistants to embodied robot agents, communication protocols should evolve to handle growing task complexity, semantic richness, and platform heterogeneity. Traditional static APIs lack the flexibility needed for such dynamic, cross-domain environments. Agents show promising potential by combining natural language understanding with structured interaction and adaptive protocol negotiation. However, developing secure and adaptive communication protocols remains a challenge to enable robust, interoperable, and intelligent agent collaboration at scale.

The design of inter‐agent protocols navigates a versatility-efficiency-portability trilemma \cite{Agora}: highly versatile protocols handle diverse tasks but incur heavy overhead, while efficient and portable APIs lack flexibility. Agents offer a compromise, using natural language and dynamic tool calls, but at the expense of latency, energy cost, and hallucination risks. 
Furthermore, as agent teams form and reconfigure dynamically, communication protocols should adapt in real time to changing roles, contexts, and capabilities. 
{Moreover, existing MCP-based ecosystems remain susceptible to tool‐poisoning attacks whereby malicious agents inject compromised plugins or APIs; decentralized A2A networks can suffer Sybil attacks from adversary-controlled identities; and Agora's reliance on natural-language routines introduces hallucination vulnerabilities when LLM-generated schemas diverge from intended semantics. Future IoA communication should embed rigorous security guarantees to build secure agent ecosystems.}

\subsection{Decentralized and Self-Governing Agent Ecosystems}\label{subsec:future2}
The shift toward decentralized IoA architectures, where smart city agents, supply chain robots, and energy grids operate without relying on a central authority, requires self-governance mechanisms that mimic biological ecosystems. Future ecosystems should enable agents to self-organize through decentralized consensus mechanisms, adaptive blockchain sharding, and bio-inspired swarm intelligence. 

However, scalable consensus protocols face a paradox \cite{9631953}: full decentralization ensures robustness but struggles with real-time coordination at scale.
Another issue is conflict resolution in emergent systems. For instance, a delivery robot may prioritize speed while a sidewalk robot emphasizes pedestrian safety. Besides, decentralized coordination may inadvertently optimize for local efficiency at the expense of global stability.

\subsection{Agent-Based Economic Systems}\label{subsec:future3}
The rise of agent-native economies calls for embedded economic frameworks within IoA architectures, such as autonomous agents trade compute resources, AI models, or sensory data. Future IoA architectures should integrate adaptive incentive mechanisms, combining short-term capacity-aware economic rewards with long-term reputation tracking, to foster trust and deter malicious participation in open agent ecosystems.

A critical challenge is preventing adversarial market manipulation in decentralized markets. Malicious agents could exploit pricing algorithms through Sybil attacks where malicious agents spawn fake identities to skew reputation scores or collude to monopolize resources. Developing Byzantine-resilient market mechanisms with verifiable fairness guarantees is essential. Additionally, cross-currency interoperability complicates transactions, as agents operating in hybrid economies (e.g., fiat-backed tokens, energy credits, reputation points) require protocol-level support for atomic swaps and trustless conversion.

\subsection{Privacy-Preserving Agent Interactions}\label{subsec:future4}
In IoA, agents increasingly handle sensitive data across domains such as healthcare and finance, where the sharing of raw data, knowledge, and inference results risks privacy breaches. While homomorphic encryption and federated learning offer partial solutions, they lack fine‑grained controls for context-aware privacy in dynamic and task‑driven environments. Future IoA architectures should integrate context-aware and task-driven privacy design, allowing agents to negotiate privacy protection levels dynamically based on task criticality, context awareness, and user consent. 

A key challenge is the latency-privacy trade-off. Techniques such as secure multi-party computation (MPC) provide strong privacy guarantees but introduce delays incompatible with real-time applications (e.g., autonomous vehicle coordination). Optimizing these trade-offs requires hardware-software co-design and lightweight cryptographic primitives. Another challenge is cross‑jurisdictional compliance. Agents operating globally should dynamically adhere to diverse regulations (e.g., GDPR and CCPA) without fragmenting collaboration.

\subsection{Cyber-Physical Secure IoA}\label{subsec:future5}
The convergence of cyber and physical layers in IoA from virtual agents to industrial robots expands attack surfaces, allowing a single compromised agent to trigger cascading physical failures. Conventional intrusion detection systems often fail to detect attacks that exploit cyber‑physical dependencies, such as LiDAR spoofing on UAVs to induce collisions. Future IoA architectures should integrate cyber-physical defenses. For instance, power‑grid agents could correlate network traffic with phasor measurement unit (PMU) readings to detect false data injection.

One major challenge lies in unified threat modeling. Bridging the semantic gap between cyber indicators (e.g., packet signatures) and physical consequences (e.g., robotic torque overloads) requires interdisciplinary frameworks that map binary-level exploits to system-level impacts. Another issue is legacy device integration. Industrial IoA agents often run on decades‑old hardware, e.g., supervisory control and data acquisition (SCADA) systems, lacking built‑in security, necessitating retrofittable, low-overhead trust anchors for secure communication with modern agents.

\subsection{Ethical \& Interoperable IoA}\label{subsec:future6}
The deployment of IoA agents in mission-critical domains, such as law enforcement, healthcare, and public policy, raises profound ethical imperatives. Ensuring ethical compliance demands transparent moral reasoning. Current black-box foundational models obscure decision logic, complicating liability assignment. Moreover, as IoA ecosystems integrate ever more heterogeneous agents, it necessites seamless interoperability on unified semantic frameworks. Future IoA systems should incorporate ethical explainability and adaptive semantic alignment. For instance, agents debate and refine value hierarchies through interaction logs and human feedback via RL \cite{cheng2024reinforcement}.

A key challenge is contextual ethics traceability. Capturing the provenance of moral decisions across chained agent interactions requires tamper-proof, explainable audit trails. Another challenge is multi-agent accountability. When collective agent behavior causes harm, how to design legal frameworks to apportion responsibility among developers, operators, and the agents themselves remain a challenge.

\section{Conclusions}\label{sec:CONSLUSION}
In this paper, we position the IoA as the next‐generation infrastructure for autonomous and interconnected intelligent systems, and have presented a comprehensive survey of IoA. Specifically, we have first provided a hierarchical IoA architecture, including its core components, distinctive features, and emerging applications. Afterward, we have explored the enabling technologies underpinning large‑scale agent collaboration, including capability discovery, dynamic task orchestration, adaptive communication protocols, consensus mechanisms, and incentive models.  
Finally, by identifying open challenges in scalability, interoperability, agent economics, security/privacy, and ethics, we have outlined a roadmap for future IoA research. As IoA continues to mature, sustained innovation in networking architectures, interoperability standards, and security paradigms will be essential to realize IoA ecosystems.

\bibliographystyle{ieeetr} 

\bibliography{ref.bib}

\end{document}